\documentclass[longauth]{aa}
\usepackage{graphicx}
\usepackage{txfonts}
\newcommand{\gsimeq}{{_>\atop^{\sim}}}

\begin{document}

   \title{The ALPINE-ALMA [CII] Survey: The nature, luminosity function, and star formation history of dusty galaxies up to z$\simeq$6}

%   \subtitle{Luminosity function of the ALPINE non-target continuum sources}
  \titlerunning{ALPINE continuum luminosity function}

   \author{C. Gruppioni
          \inst{1} \and M. Bethermin  \inst{2} \and F. Loiacono \inst{3,1} \and O. Le F\'evre \inst{2} \and P. Capak \inst{4} \and P. Cassata  \inst{5,6} \and 
A.L. Faisst \inst{4} \and D. Schaerer \inst{7} \and J. Silverman \inst{8} \and L. Yan \inst{9} \and S. Bardelli \inst{1} \and M. Boquien \inst{10} \and R. Carraro \inst{11} \and  A. Cimatti \inst{3,12} \and M. Dessauges-Zavadsky \inst{7} \and 
        M. Ginolfi \inst{7} \and  S. Fujimoto \inst{13,14} \and N.P. Hathi  \inst{15} \and  G.C. Jones  \inst{16,17} \and  Y. Khusanova  \inst{2} \and A.M. Koekemoer \inst{15} \and G. Lagache \inst{2} \and B.C. Lemaux \inst{18} \and
                P.A. Oesch \inst{7,13} \and F. Pozzi \inst{3} \and D.A. Riechers \inst{19,20} \and G. Rodighiero \inst{5} \and M. Romano \inst{5,6} \and  M. Talia \inst{1,3} \and L. Vallini \inst{21} \and D. Vergani \inst{1} \and G. Zamorani \inst{1} \and E. Zucca \inst{1}             
         }
   \institute{Istituto Nazionale di Astrofisica: Osservatorio di Astrofisica e Scienza dello Spazio di Bologna, via Gobetti 93/3, 40129, Bologna, Italy;
              \email{carlotta.gruppioni@inaf.it} 
            \and 
            Aix Marseille University, CNRS, LAM, Laboratoire d'Astrophysique de Marseille, Marseille, France
         \and
            Dipartimento di Fisica e Astronomia, Universit\`a of Bologna, via Gobetti 93/2, 40129, Bologna, Italy
           \and
          Infrared Processing and Analysis Center, California Institute of Technology, Pasadena, CA 91125, USA 
          \and
          Dipartimento di Fisica e Astronomia, Universit\`a di Padova, vicolo Osservatorio 3, 35122 Padova, Italy  
          \and
          Istituto Nazionale di Astrofisica: Osservatorio Astronomico di Padova, Vicolo dell'Osservatorio 5, I--35122, Padova, Italy
          \and
          Observatoire de Gen\`eve, Universit\'e de Gen\`eve, 51 Ch. des Maillettes, 1290 Versoix, Switzerland
         \and
         Kavli Institute for the Physics and Mathematics of the Universe, The University of Tokyo, 5-1-5 Kashiwanoha, Kashiwa-shi, Chiba, 277-8583 Japan       
           \and
The Caltech Optical Observatories, California Institute of Technology, Pasadena, CA 91125, USA
\and
         Centro de Astronom\'ia (CITEVA), Universidad de Antofagasta, Avenida Angamos 601, Antofagasta, Chile
 \and
Instituto de F\'isica y Astronom\'ia, Universidad de Valpara\'iso, Gran Breta$\tilde{\rm n}$a 1111, Playa Ancha, Valpara\'iso, Chile
\and
INAF - Osservatorio Astrofisico di Arcetri, Largo E. Fermi 5, I-50125, Firenze, Italy
\and
Cosmic Dawn Center (DAWN), Copenhagen, Denmark
\and
Niels Bohr Institute, University of Copenhagen, Lyngbyvej 2, DK2100 Copenhagen, Denmark
\and
Space Telescope Science Institute, 3700 San Martin Drive, Baltimore, MD 21218, USA
 \and
  Cavendish Laboratory, University of Cambridge, 19 J. J. Thomson Ave., Cambridge CB3 0HE, UK
 \and
Kavli Institute for Cosmology, University of Cambridge, Madingley Road, Cambridge CB3 0HA, UK
\and 
 Department of Physics, University of California, Davis, One Shields Ave., Davis, CA 95616, USA
 \and
Department of Astronomy, Cornell University, Space Sciences Building, Ithaca, NY 14853, USA
\and
Max-Planck-Institut f\"ur Astronomie, K\"onigstuhl 17, D-69117 Heidelberg, Germany
\and
Leiden Observatory, P.O. Box 9513 NL-2300 RA, NL           
          %            \email{c.ptolemy@hipparch.uheaven.space}
  %           \thanks{The university of heaven temporarily does not
   %                  accept e-mails}
             }

   \date{Received 25 May, 2020 / Accepted 14 September, 2020}

% \abstract{}{}{}{}{} 
% 5 {} token are mandatory
  \abstract
   {}
   {We present the detailed characterisation of a sample of 56 sources serendipitously detected in ALMA band 7 as part of the ALMA Large Program to 
   INvestigate CII at Early Times (ALPINE). These sources, detected in COSMOS and ECDFS, have been used to derive the total infrared luminosity function (LF)
   and to estimate the cosmic star formation rate density (SFRD) up to $z$$\simeq$6.
}
   {We looked for counterparts of the ALMA sources in all the available multi-wavelength (from HST to VLA) and photometric redshift catalogues. 
   We also made use of deeper UltraVISTA and {\em Spitzer} source lists and maps to identify optically dark sources with no matches 
   in the public catalogues. We used the sources with estimated redshifts to derive the 250-$\mu$m rest-frame and total infrared 
   (8--1000\,$\mu$m) LFs from $z$$\simeq$0.5 to 6.
}
   {Our ALMA blind survey (860-$\mu$m flux density range: $\sim$0.3--12.5 mJy) allows us to further push the study of 
   the nature and evolution of dusty galaxies at high-$z$, identifying luminous and massive sources to redshifts and faint luminosities never 
   probed before by any far-infrared (far-IR) surveys. The ALPINE data are the first ones to sample the faint end of the infrared LF, 
   showing little evolution from $z$$\simeq$2.5 to $z$$\simeq$6, and a `flat' slope up to the highest redshifts (i.e. 4.5$<$$z$$<$6). 
   The SFRD obtained by integrating the luminosity function remains almost constant between $z$$\simeq$2 and $z$$\simeq$6, 
   and significantly higher than the optical or ultra-violet (UV) derivations, showing a significant contribution of dusty galaxies and obscured star formation at high-$z$. 
   About 14\% of all the ALPINE serendipitous continuum sources %(but 75\% of the sample without counterparts in the photometric redshift catalogues)
   are found to be optically and near-infrared (near-IR) dark (to a depth Ks$\sim$24.9 mag). Six show a counterpart only in the mid-IR and no HST or near-IR identification,
   while two are detected as [C~II] emitters at $z$$\simeq$5. The six HST$+$near-IR dark galaxies with mid-IR counterparts are found to contribute about 
   17\% of the total SFRD at $z$$\simeq$5 and to dominate the high-mass end of the stellar mass function at $z$$>$3.
   }
  % conclusions heading (optional)
  % leave it empty if necessary 
  { }
   \keywords{galaxies: evolution -- galaxies: high-redshift -- galaxies: luminosity function -- cosmology: observations -- submillimeter: galaxies}
  \maketitle
%
%-------------------------------------------------------------------
%----------------------------------------------------------------- 
\section{Introduction}
\label{sec:intro}
Our current knowledge of the cosmic star formation rate density (SFRD) at high redshift ($z$$>$3) is 
based mostly on galaxy samples selected in the ultra-violet (UV) rest frame (e.g. \citealt{bouwens15} and \citealt{oesch18}),
whose bolometric star formation rates (SFRs) are not measured, but rather inferred through uncertain dust-correction techniques,
and which are not necessarily representative of the whole galaxy population (e.g. missing strongly obscured massive systems with high dust content). 

Since the discovery of the cosmic infrared background (CIB; representing the cumulative emission reprocessed by dust from all the
galaxies throughout the cosmic history of the Universe; e.g. \citealt{lagache05}) at the end of the 1990s by the COBE satellite (\citealt{puget96,hauser98}),
and its resolution into discrete, rapidly evolving, far-infrared (far-IR) and sub-millimetre (sub-mm) sources 
 by deep extragalactic surveys performed with the Infrared Space Observatory (ISO) and the Submillimetre Common-User Bolometer Array (SCUBA) on the 
James Clerk Maxwell Telescope (JCMT), 
many searches have focused on deriving how much star formation activity in the early Universe is obscured by dust.
These dusty star-forming galaxies, also called 'sub-millimetre galaxies' (SMGs; e.g. \citealt{smail97} and \citealt{hughes98}; \citealt{barger98}; \citealt{blain02}), 
are characterised by large far-IR luminosities ($\gtrsim$10$^{12}$ L$_\odot$) and stellar masses (M$\gtrsim$7$\times$10$^{10}$ M$_\odot$; e.g. \citealt{chapman05}, \citealt{simpson14}), 
extremely high star formation rates (SFRs, $\gtrsim$100 M$_\odot$ year$^{-1}$; e.g. \citealt{swinbank14}), and large gas reservoirs
($\gtrsim$10$^{10}$ M$_\odot$; e.g. \citealt{bothwell13}, \citealt{bethermin15}).
Despite them being rare and luminous objects, typically located around $z$$\sim$2--2.5 (e.g. \citealt{chapman03}, \citealt{wardlow11}), 
their tremendous SFRs make them substantial contributors to the SFRD at Cosmic Noon, that is, 1$<$$z$$<$3 (e.g. \citealt{casey13}).
However, the fraction of dust-obscured star formation, which is traced by {\em Spitzer} and AKARI up to $z$$\simeq$2 (e.g. \citealt{rodighiero10}, \citealt{goto19}),
and by {\em Herschel} up to $z$$\simeq$3 (e.g. \citealt{gruppioni13}, \citealt{magnelli13}),
is still unknown at higher redshifts.

One of the problems is the difficulty in identifying the SMGs because of the coarse
angular resolution of single-dish telescopes and the faintness of the optical/UV counterparts. 
The few SMGs that have been identified at $z$$>$4 trace only the bright tail
of the SFR distribution (e.g. \citealt{capak11, walter12, riechers11, riechers13, riechers17, marrone18}) 
and are unlikely to represent the bulk of the population.
Moreover, most of the SMGs have photometric or spectroscopic observations that likely place them at $z$$<$3 (\citealt{brisbin17}). 

The Atacama Large Millimetre/submillimetre Array (ALMA) has now opened a gap in the wall, allowing us to
refine our understanding of dusty galaxies at high redshifts by unveiling less extreme galaxies, 
between massive SMGs and normal star-forming galaxies, through superb sensitivity and high spatial 
resolution surveys in the sub-mm/mm domain. 
This can be achieved thanks to the recently explored ability of ALMA to reveal serendipitously detected galaxies in
blind extragalactic surveys.

The ALMA deep surveys performed by \citet{dunlop17}, \citet{walter16}, and \citet{aravena16}, probing very faint fluxes
over small areas ($<$5 arcmin$^2$), as well as the wider (covering few tens of arcmin$^2$) and shallower (to $\sim$100--200 $\mu$Jy) 
surveys by \citet{hatsukade18} and \citet{franco18}, have enabled us to uncover faint (sub-)mm 
populations at $z$$>$4, with infrared luminosities ($L_{IR}$, between 8 and 1000 $\mu$m) $\lesssim$10$^{12}$ L$_\odot$ (e.g. \citealt{yamaguchi19}).
An important product of these surveys is the discovery of a population of faint ALMA galaxies that are undetected even in the deepest optical and 
near-infrared (near-IR; i.e. $\simeq$1--3 $\mu$m) images with the Hubble Space Telescope (HST). These 'HST-dark' galaxies are often identified in the mid-infrared (mid-IR),
in deep {\em Spitzer}-IRAC 3.6 or 4.5-$\mu$m images (e.g. \citealt{franco18,wang19,yamaguchi19}), although, despite them being unlikely 
spurious ALMA detections (e.g. \citealt{williams19, romano20}), some remain undetected even in IRAC maps. 

Indeed, ALMA follow-up studies of 'classical' SMGs (e.g. \citealt{simpson14, dudzeviciute20}) have found that a fraction of these objects are invisible in deep optical/near-IR images;
in particular, \citet{dudzeviciute20} found that $\sim$17\% of the SCUBA-2 SMGs in the UKIDSS/UDS field (AS2UDS survey) are undetected in the optical/near-IR down to Ks$=$25.7 mag. 
Already in the pre-ALMA studies there were cases of bright SMGs, such as the source GOODS 850-5 (also known as GN10), undetected in very deep optical/near-IR images (e.g. \citealt{wang09}),
while the existence of dusty star-forming systems at $z$$>$5 was already discussed by \citet{dey99} in the late 90s as a realistic expectation.

The HST-dark galaxies also tend to be serendipitously found in CO line scan surveys (see e.g. \citealt{riechers20}, who found two of them at $z$$>$5), possibly with space 
densities higher than expected even at the bright end of the CO luminosity functions (LFs). 
These results indicate the existence of a prominent population of dusty star-forming
galaxies at $z$$>$4, which is fainter than the confusion limit of the single-dish sub-mm surveys that discovered the SMGs, but with much larger space densities, providing 
a significant contribution to the SFRD at high-$z$, even higher than that of the UV-bright galaxies at the same redshifts (e.g. \citealt{rodighiero07,williams19,wang19}).

Very faint ALMA fluxes were also reached by surveys of serendipitously detected sources in targeted observations (i.e. non-pure-blind surveys), which were able to
constrain the faint end of the sub-mm/mm galaxy source counts, estimate their contribution to extragalactic background light, study their nature, and possibly detect
dark galaxies (e.g. \citealt{hatsukade13,ono14,carniani15,oteo16, fujimoto16}).

Here, we present the identification, multi-wavelength characterisation, and luminosity function of a sample of 56 sources, serendipitously detected in
continuum at $\sim$860 and $\sim$1000 $\mu$m (ALMA band 7), within the {\em ALMA Large Program to INvestigate CII at Early Times} (ALPINE, PI: LeF\'evre; see 
\citealt{lefevre19, faisst20, bethermin20})\footnote{https://cesam.lam.fr/a2c2s/} survey fields. Firstly, ALPINE is a 70-hour ALMA survey in band 7, specifically designed to measure singly ionised 
carbon ([C~II] at 158~$\mu$m) 
emission and any associated far-IR continuum for 118 main-sequence galaxies at 4.4$<$$z$$<$5.9 (representative in stellar mass and SFR of the star-forming population at $z$$\simeq$5; see \citealt{lefevre19,faisst20}). 
The programme, completed in February 2019, allows us to build a coherent picture of the baryon cycle in galaxies at $z$$>$4 for the first time, 
by connecting the internal ISM properties to their well-characterised stellar masses and SFRs (from a wealth of ancillary photometric and spectroscopic data, which is already in hand). 
All the ALPINE pointings are
located in the Extended Chandra Deep Field South (ECDFS; \citealt{giacconi02})
and Cosmic Evolution Survey (COSMOS; \citealt{scoville07}), thus benefitting from a
wealth of ancillary multi-wavelength photometric data (from UV
to far-IR), making ALPINE currently one of the largest panchromatic
samples to study the physical properties of normal 
galaxies at high-$z$.

Besides the main targets, in the ALPINE pointings a blind search for serendipitous line and/or continuum emitters has been performed in a total area of 24.9 arcmin,$^2$ providing
two independent catalogues of emission-line (\citealt{loiacono20}) and continuum (\citealt{bethermin20}) detections.
This work focuses on the continuum sample of serendipitous detections.
For these sources, we performed identifications in all the catalogues and deep images available in the COSMOS and ECDFS fields. We
also constructed spectral energy distributions (SEDs), estimated photometric redshifts when they were unavailable in the literature, and derived the 250-$\mu$m rest-frame
and total IR (8--1000 $\mu$m) luminosity functions and the contribution of dusty galaxies to the cosmic SFRD up to $z$$\simeq$6.

The ALPINE sample of serendipitous continuum galaxies is briefly described in Section~\ref{sec:det}, the identification process and results are presented in 
Section~\ref{sec:nature}, while the luminosity function results are discussed in Section~\ref{sec:lf}. In Section~\ref{sec:SFRD}, we derive the contribution of our sources to the 
cosmic SFRD, and in Section~\ref{sec:concl} we present our conclusions.
Throughout the paper, we use a \citet{chabrier03} stellar initial mass function (IMF) and adopt a $\Lambda$CDM cosmology
with H$_0$=70 km s$^{-1}$ Mpc$^{-1}$, $\Omega_m$=0.3, and $\Omega_\Lambda$=0.7.
%--------------------------------------------------------------------
%----------------------------------------------------------------- 
\section{The ALPINE serendipitous continuum detections}
\label{sec:det}

The ALMA ALPINE observations were carried out in band 7 during Cycle 5 and Cycle 6, and were completed in February 2019. 
Each target was observed for $\sim$30 minutes of on-source integration time, with the phase centre pointed at the UV position of the sources. 
One spectral window was centred on the [C~II] expected frequencies, according to the spectroscopic redshifts extracted from the UV spectra, while 
the other side bands were used for continuum measurements only. 
The data were calibrated using the Common Astronomy Software Applications package (CASA; \citealt{mcmullin07}), version 5.4.0, and 
the continuum maps were obtained by collapsing the line-free channels in all the spectral windows (see \citealt{bethermin20}). 
%The typical angular resolution of the final products, computed as average beam axis, is 0.9$^{"}$ (between 5.2 and 6 kpc at 4.4$<$$z$$<$6), with values 
%in the range 0.8$^{"}$--1$^{"}$. 

The ALMA observational strategy/setup, the details of the data reduction and the method adopted to extract continuum 
flux density information from ALPINE data and to select a complete sample of serendipitous sources, are comprehensively discussed in \citet{bethermin20}. 
In the following paragraphs we summarise the main steps.
The data cubes were imaged using the {\tt tclean} CASA routine down to a flux threshold of 3$\sigma$ ($\sigma$ being the standard deviation measured in
a non-primary-beam-corrected map after masking the sources). A natural weighting of the visibilities was applied in order to maximise the point-source
sensitivity and to optimise the measurement of the integrated properties of the ALPINE targets.
The continuum maps were obtained by excluding the channels contaminated target source lines and those of a few
off-centre, serendipitously detected continuum sources with lines. In fact, in order to avoid possible contamination of the continuum flux 
by line flux, spectra were extracted for all the non-target sources and new tailored continuum maps were produced by masking the potentially 
line-contaminated channels and then remeasuring the continuum flux (correction varying from 58\% to a negligible fraction of the flux density). 

The average synthesized beam size is 1.13$\times$0.85 arcsec$^2$ (size varies with frequency and array configuration, i.e. between 5.2 and 6 kpc at 4.4$<$$z$$<$6).
The continuum sensitivity also varies with the frequency, and for this reason the continuum sources have been extracted on %non-primary-beam corrected 
signal-to-noise ratio (SNR) maps, 
by searching for local maxima above a given threshold using the {\tt find\_peak} routine of {\tt astropy}. 
As revealed from simulations shown in \citet{bethermin20}, the threshold above which we obtain a purity 
of 95\% corresponds to an SNR$=$5 outside the central
region of 1 arcsec radius (expected to contain the ALMA continuum flux of the ALPINE targets). The sources extracted in the 1-arcsec
central regions are referred to as `target', and the objects found outside of this area are referred to as `non-target'. In this paper, we focus only on the non-target sources.

The final sample of non-target sources detected in continuum at SNR$>$5 in ALMA band 7 consists of 56 sources, 
of which three are in the ECDFS and 53 are in COSMOS, and which were extracted over a total area of 24.92 arcmin$^2$ (excluding the circle of 1 arcsec radius around the central ALPINE targets). 
The number of expected spurious sources in this sample is $\leq$3, while the completeness is a function of the flux density and the size of each source (see \citealt{bethermin20}),
as discussed in Section~\ref{sec:lf}.
One of the ECDFS sources has been detected in two different (slightly overlapping) ALPINE pointings, therefore it has a flux measurement in both channels, that is,
860~$\mu$m and 1000~$\mu$m. Details on the flux measurement and uncertainties are provided in \citet{bethermin20}.
%-----------------------------------------------------------------
%----------------------------------------------------------------- 
\section{The nature of the ALPINE serendipitous sources}
\label{sec:nature}
We took advantage of the great wealth of multi-wavelength ancillary data, catalogues, spectroscopic and photometric redshifts, and deep images that are available 
in the ALPINE fields (ECDFS and COSMOS; see e.g. \citealt{faisst20}) to investigate the nature of the serendipitous sources detected in continuum by ALMA.

The ground-based photometry available in the ECDFS includes $U38$, $b$, $v$, $Rc$, and $I$ broad-band filters from the Wide Field Imager on the ESO/2.2-m 
telescope; $U$ and $R$ bands from VIMOS on the ESO VLT; near-IR filters $J$, $H$, and $Ks$ from ISAAC on the ESO VLT; $J$ and $Ks$ data from WIRCam on the CFHT;
and 14 intermediate-band filters from the Suprime-Cam on the Subaru telescope. In addition, a wealth of HST observations are available in the ECDFS field. 

The photometric data available in the COSMOS field include $u$-band observations from MegaCam on CFHT;  $B$, $V$, $r$+, $i$+, $z$++, as
well as 12 intermediate-band and two narrow-band filters from the Suprime-Cam on Subaru; $YHSC$-band from
the Hyper Suprime-Cam on Subaru; as well as near-IR bands $H$ and $Ks$ from WIRCam on CFHT; and $Y$ , $J$,
$H$, and $Ks$ from VIRCAM on the ESO-VISTA telescope. In terms of HST data, all but one ALPINE pointing in COSMOS are covered by ACS
$F814W$ observations (\citealt{koekemoer07, scoville07}). This is also the case for CANDELS data in ACS and WFC3 bands (\citealt{grogin11, koekemoer11}) and several additional 
pointings in ACS and WFC3 bands.
%--------------------------------------------------------------------   
   \begin{figure*}
   \centering
   \includegraphics[width=17cm]{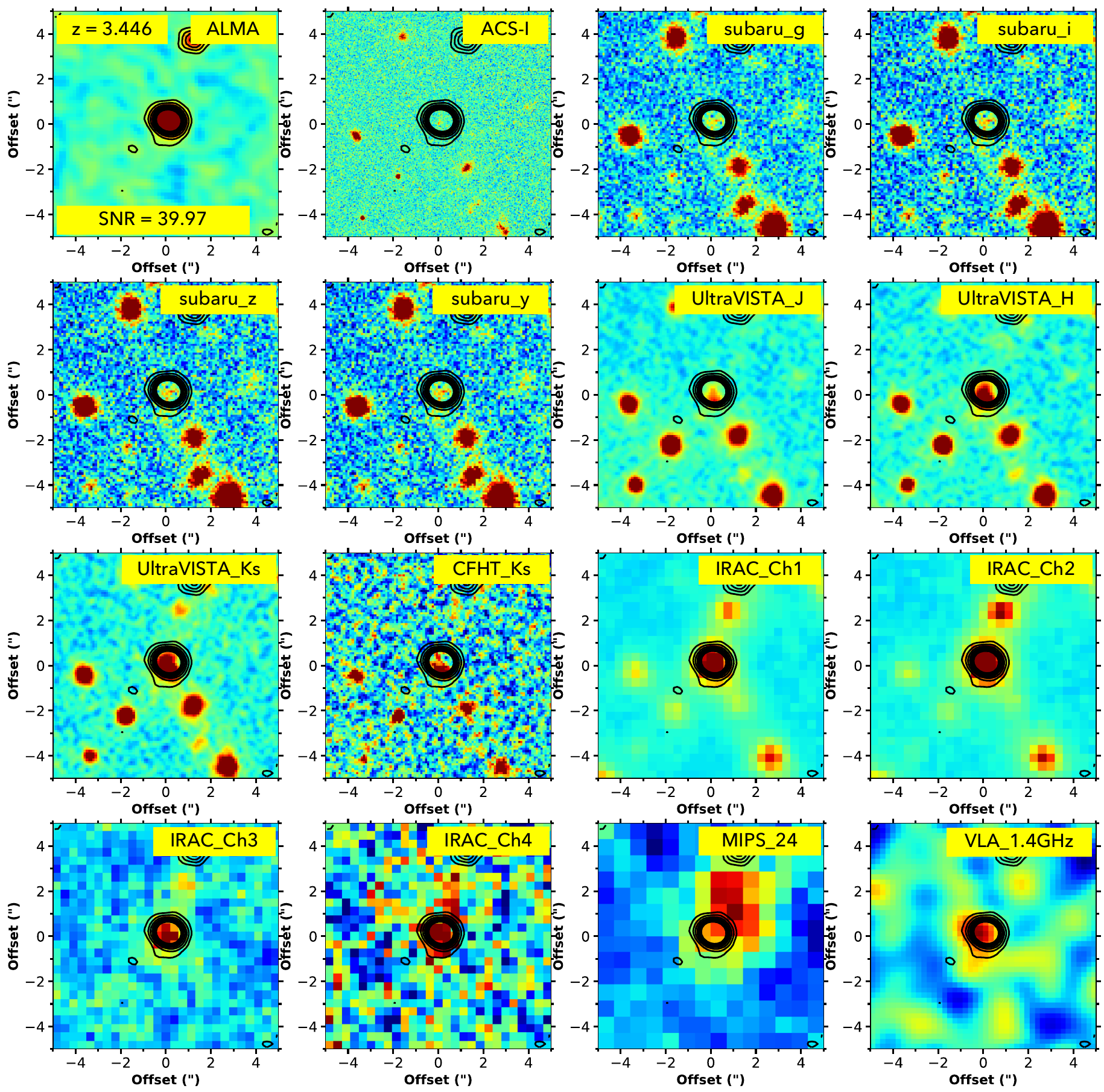}
   \caption{Example of identification of ALPINE non-target continuum source: the postage stamps (from $top$ $left$ to $bottom$ $right$) show the ALMA band 7 continuum map and the 
   ALMA $\geq3$$\sigma$ contours over-plotted on images from HST/ACS-i to radio VLA-1.4GHz (band specified in the top-right corner). 
   $(a)$ - Object with multi-wavelength counterparts in all bands and photo-$z$ from \citet{laigle16}. The plotted contour levels are at 3, 5, 7, 9, 11, 13, and 15$\sigma$, corresponding to 
   0.32, 0.53, 0.74, 0.95, 1.16, 1.37, and 1.58 mJy. The ALMA 1-mm flux density of this source (SC$\_$1$\_$DEIMOS$\_$COSMOS$\_$910650) is  4.22 mJy, the RMS (1$\sigma$) of this pointing is $\sim$0.11 mJy. Continued on the next page. }
              \label{Fig:optID} %
    \end{figure*}
%--------------------------------------------------------------------
   \begin{figure*}
  \centering
   \includegraphics[width=17cm]{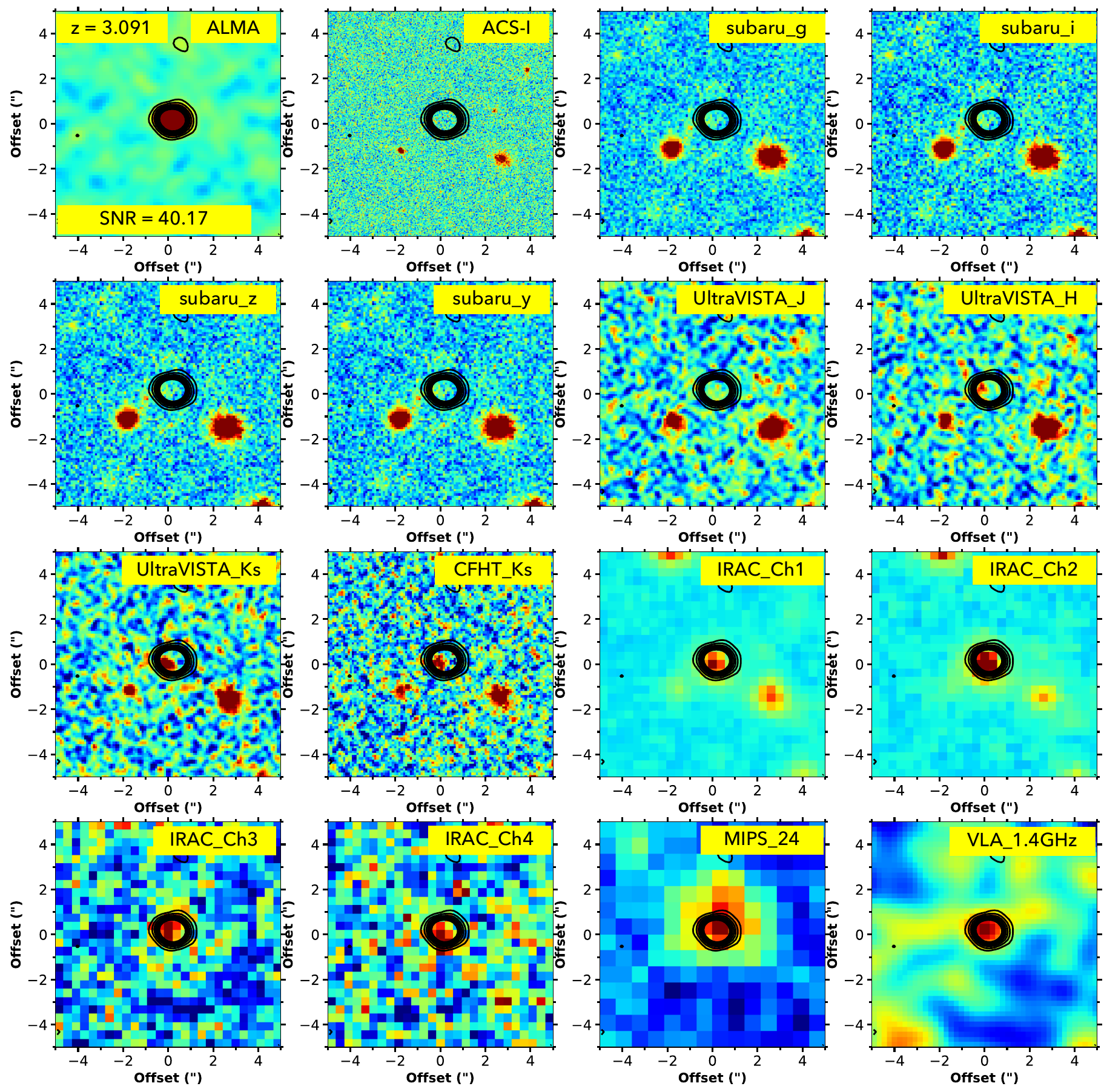}
   \caption{Object with no optical counterpart, but with multi-wavelength counterparts from near-IR to sub-mm and photo-$z$ (from \citealt{laigle16}). The plotted contour levels are at 3, 5, 7, 9, 11, 13, and 15$\sigma$, corresponding to 
   0.26, 0.44, 0.62, 0.79, 0.97, 1.15, and 1.32 mJy. The ALMA 860-$\mu$m flux density of this source (SC$\_$1$\_$DEIMOS$\_$COSMOS$\_$680104) is 3.54 mJy, the RMS (1$\sigma$) of this pointing $\sim$0.088 mJy.} 
              \label{Fig:optID2}
                  \end{figure*}
%--------------------------------------------------------------------
   \begin{figure*}
   \centering
   \includegraphics[width=17cm]{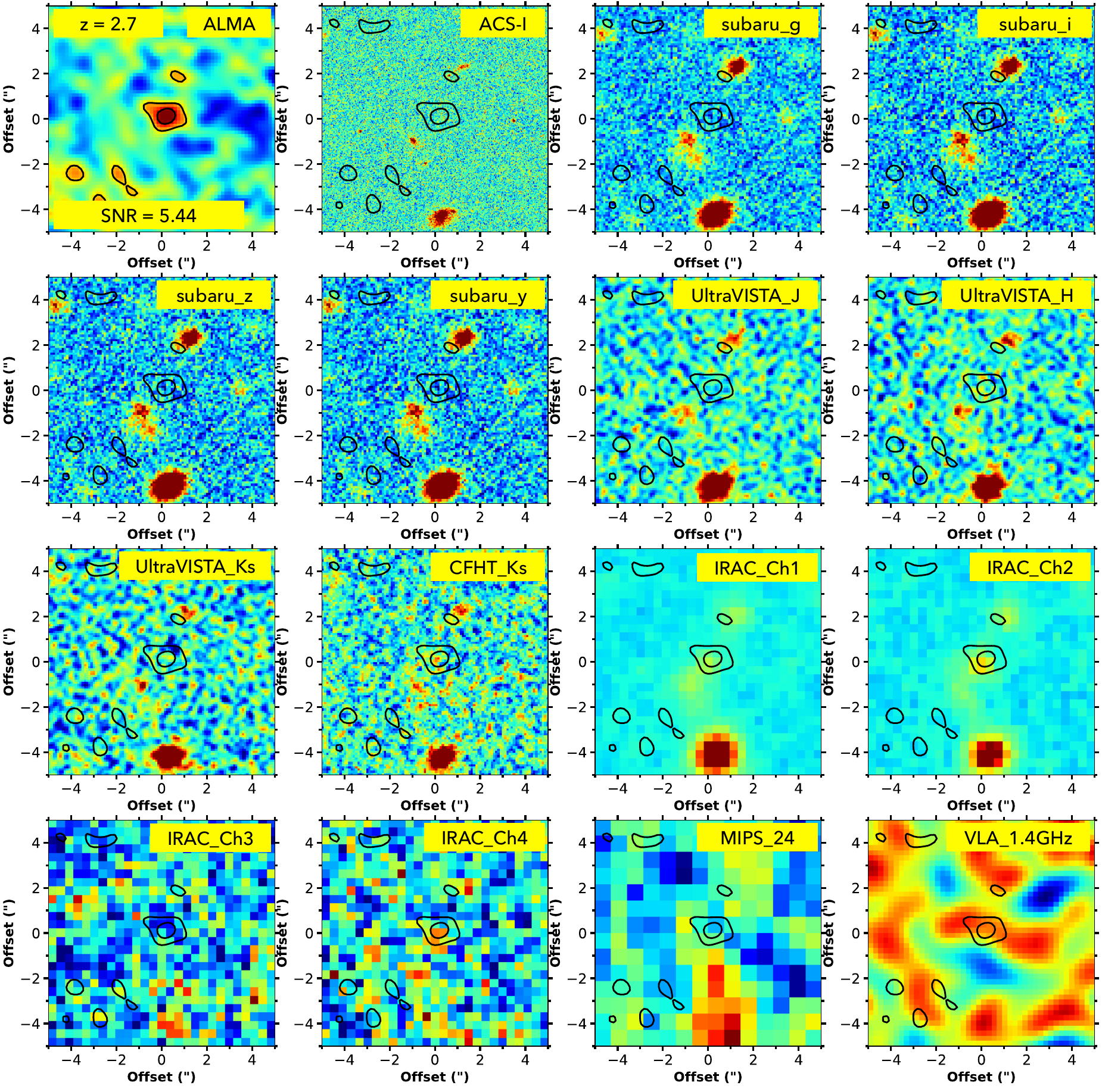}
   \caption{Optically dark galaxy detected only in deep IRAC-SPLASH 4.5-$\mu$m catalogue. The photo-$z$ has been derived with \textsc{Le Phare}
   using ALMA and IRAC data only. The plotted contour levels are at 3, 5$\sigma$, corresponding to 
   0.14, 0.23 mJy. The ALMA 1-mm flux density of this source (SC$\_$1$\_$DEIMOS$\_$COSMOS$\_$224751) is  0.25 mJy, the RMS (1$\sigma$) of this pointing is $\sim$0.046 mJy.}
              \label{Fig:optID3}
                  \end{figure*}
%--------------------------------------------------------------------
   \begin{figure*}
   \centering
   \includegraphics[width=17cm]{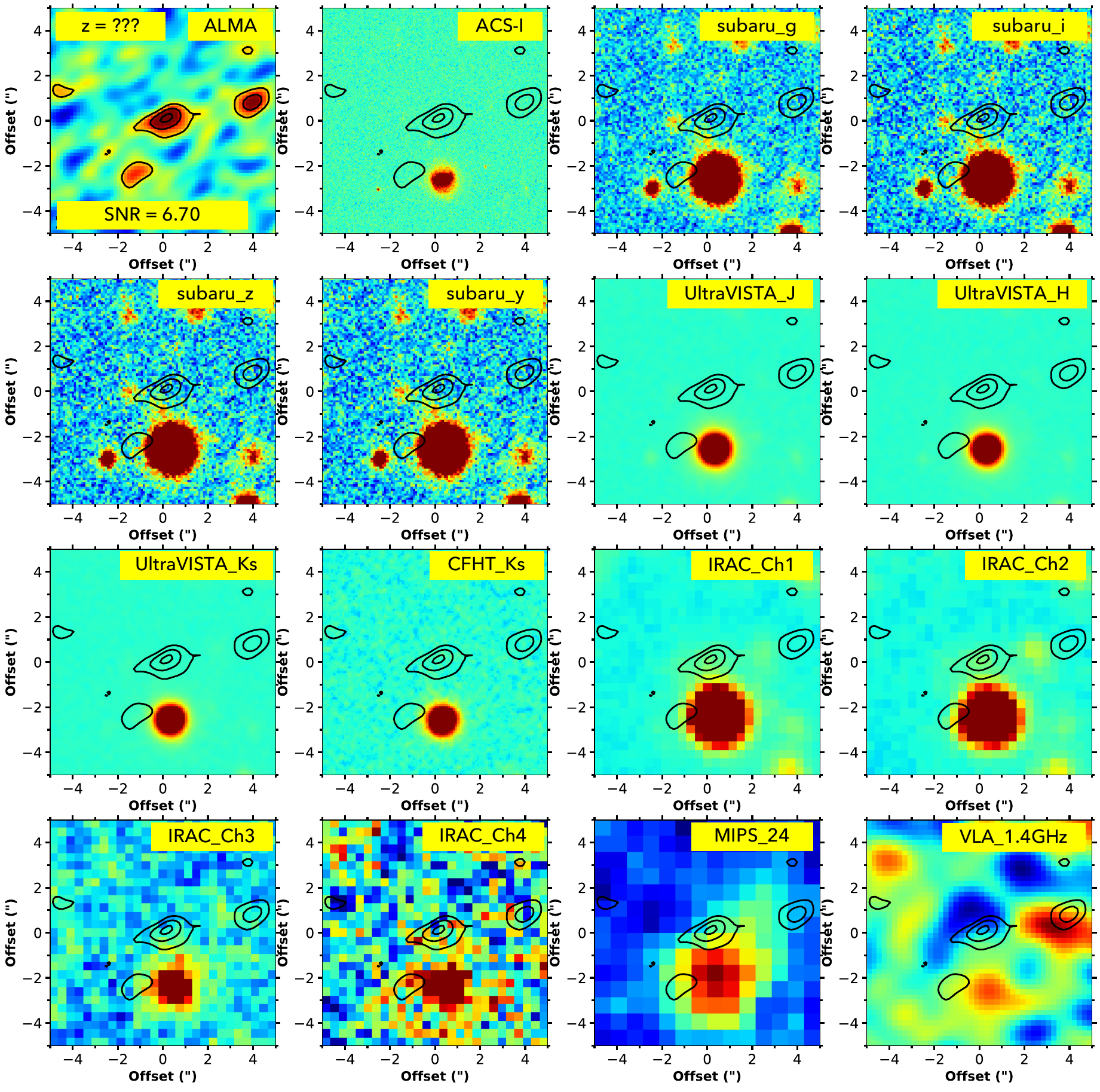}
   \caption{ALPINE serendipitous source with no obvious identification in any bands. The plotted contour levels are at 3, 5, and 6.5$\sigma$, corresponding to 
    0.47, 0.79, and 1.02 mJy. The ALMA 1-mm flux density of this source (SC$\_$1$\_$DEIMOS$\_$COSMOS$\_$471063) is 1.06 mJy, the RMS (1$\sigma$) of this pointing is $\sim$0.16 mJy.}
              \label{Fig:optID4}
    \end{figure*}
%----------------------------------------------------------------- 
The space-based photometry in both fields includes {\em Spitzer} data in the four IRAC bands (3.6, 4.5, 5.8, and 8.0 $\mu$m)
and in the MIPS 24-$\mu$m band, and Herschel data in the PACS (100 and 160 $\mu$m) and SPIRE bands (250, 350, and 500 $\mu$m).
A detailed summary and references of the different ground- and space-based data available in the two fields are presented in \citet{faisst20}.

In the identification process, we first matched the ALMA non-target list with the 3D-HST catalogues (\citealt{brammer12, skelton14, momcheva16}) in 
both ECDFS and COSMOS, and with the COSMOS2015 (\citealt{laigle16}) in COSMOS. Then, for the COSMOS sources, we looked for counterparts in the the super-deblended (\citealt{jin18}) and the DR4 UltraVISTA catalogues 
(\citealt{mccracken12}; Moneti et al. 2019\footnote{http://www.eso.org/rm/api/v1/public/releaseDescriptions/132}). 
Moreover, in COSMOS we considered the IRAC catalogue based on the Spitzer Large Area Survey with Hyper-Suprime-Cam data 
(SPLASH; \citealt{capak12, steinhardt14}).\footnote{The SPLASH maps are available, upon request, at http://splash.caltech.edu/}
In the following sections, we describe in detail the identification process of the ALPINE non-target continuum sources and the results obtained.
%----------------------------------------------------------------- 
%----------------------------------------------------------------- 
\subsection{Source identification}
\label{sec:id}
\subsubsection{Catalogue match}
\label{sec:match}
As a first step in the identification process of the ALPINE non-target sources, we cross-matched the ALMA list with the multi-wavelength catalogues available from the
literature in COSMOS and ECDFS. We found a counterpart within 1 arcsec of the source position for all three GOODS-S galaxies in the 3D-HST catalogue, and 
for 38 of the COSMOS sources in the \citet{laigle16} catalogue. We then found one additional COSMOS source in the 3D-HST catalogue (three in total, but two in common with \citealt{laigle16}), while other three (39 in total, but 36 already in the COSMOS2015 catalogue)
were identified with galaxies in the super-deblended catalogue (at $\lambda$$\geq$24 $\mu$m, as well as two also in the UltraVISTA DR4 catalogue) by \citet{jin18}, plus another three only with the IRAC/SPLASH objects (the fluxes were provided by M. Giulietti, private communication).
By running Monte Carlo random shifts of the COSMOS15 catalogue, we find an average number of spurious matches $\lesssim$2, at an average distance of $\sim$0.7$^{\prime \prime}$ from the ALMA sources.
Since the great majority of the positional offsets between the ALMA sources and the catalogue counterparts are $<$0.4$^{\prime \prime}$, with just six sources with an offset in the range of 0.4$^{\prime \prime}$--0.6$^{\prime \prime}$, we can consider that the false match rate
as negligible.
Moreover, as we discuss later, we visually inspected the ALMA contours over-plotted onto the images in all the available bands and for all the sources, in order to validate the match.
We were therefore able to photometrically identify 48 sources out of 56 (3/3 in ECDFS and 45/53 in COSMOS), leaving us with a sample of eight galaxies with no counterpart in any of the available catalogues. 
Of these eight sources, two were identified as line emitters in the blind-lines catalogue by \citet{loiacono20}.
Because the two serendipitously detected lines associated with unidentified continuum sources are in the same side band as the [C~II]~158$\mu$m emission of the ALPINE targets in the same pointings, they are likely [C~II] 
as well (see e.g. \citealt{jones20,romano20}). This provided us with a spectroscopic redshift estimate for two sources without any catalogue counterparts, leaving us with six sources with neither catalogue matches nor redshift estimates.  
%----------------------------------------------------------------- 
%----------------------------------------------------------------- 
\subsubsection{Image visual inspection}
\label{sec:image}
In a second step, we inspected the images - from UV to sub-mm and radio - at the position of the ALMA sources, finding a likely faint counterpart
(i.e. below the threshold imposed by the catalogues, at 2.5--4.5$\sigma$) in the IRAC/SPLASH maps (at 4.5 $\mu$m) for two of the unidentified sources. % and in the MIPS 24-$\mu$m image for 1.
By inspecting the images, we also found two sources for which the optical counterpart from \citet{laigle16} - though within 1 arcsec from the ALMA position - was slightly offset and likely not the true
identification. This is because at longer wavelengths (i.e. Ks and IRAC bands) another source appeared at the exact position of the ALMA galaxy. 
For these sources, only the long wavelength photometric data ($\gtrsim$2 $\mu$m, assumed to represent the true identification) were considered for constructing the spectral energy distribution (SED).
\begin{table*}
\caption{Summary of continuum source identification}             % title of Table
\label{tab:ID}      % is used to refer this table in the text
\centering                          % used for centering table
\begin{tabular}{| c | c | c c c c c c c |}        % centered columns (4 columns)
\hline\hline               % inserts double horizontal lines
   Redshift & \multicolumn{8}{c}{Photometry} \vline \\ \hline 
      & TOT &   COSMOS2015  & 3D-HST\tablefootmark{a} & UVDR4 & SPLASH & Super-deblended & Ad-hoc IRAC  & No ID  \\  % table heading 
\hline    
                  % inserts single horizontal line
   TOT   & 56 & 38  &  3$+$1 (3$+$3)\tablefootmark{b}  &  2 (26)\tablefootmark{c} &  3 (42) & 1 (39) &  2 & 6 \\    \hline  % inserting body of the table
    Catalogue  &  38 &  33 &  4 & 0  & 0  &  1 & 0   & 0   \\
    \textsc{Le Phare} & 9 & 2 & 0 & 2 & 3 & 0 & 2 & 0 \\ 
    
    [C~II] & 5 & 3 & 0 & 0 & 0 & 0 & 0  & 2 \\ 
    No $z$ & 4  &  0   & 0 & 0  & 0 &  0 & 0 &4 \\
\hline     \hline                             %inserts single line
\end{tabular}
\tablefoot{
\tablefoottext{a}{ECDFS$+$COSMOS.}\\
\tablefoottext{b}{Values outside parentheses are the 'new' identifications not included in other catalogues, while those in parentheses are the total number of sources  identified in that catalogue, some of which are already included in other catalogues: for example, COSMOS2015.}\\
\tablefoottext{c}{Twenty-four of the 26 galaxies found in the new UltraVISTA DR4 catalogue are also in COSMOS2015, while two are in the super-deblended list. }}
\end{table*}

In the end, the number of sources with no obvious identification (either photometric nor spectroscopic) is four, which is more or less consistent with the number of expected spurious 
detections estimated through inverted map analysis: 2.8$^{+2.9}_{-1.6}$ (see \citealt{bethermin20}). The SNRs of these four unidentified sources are 9.3, 6.7, 5.5, and 5.1:
while the latter two are likely spurious detections, for the other two, this conclusion is not so obvious, since they were detected at a significantly high SNR. They could be dark galaxies with a mid-IR counterpart just below the detection threshold.
To summarise, among the 56 continuum sources, 44 were identified in the optical and/or near-IR (38 COSMOS2015, four 3D-HST, two UltraVISTA DR4), six only in the mid-IR (three SPLASH, one super-deblended, two IRAC images), two in [C~II] line (with no photometric counterpart), 
and four remained unidentified (two of which could be spurious).
The results of our identification process are summarised in Table~\ref{tab:ID}.
%----------------------------------------------------------------- 
   \begin{figure}
   \centering
   \includegraphics[width=9cm]{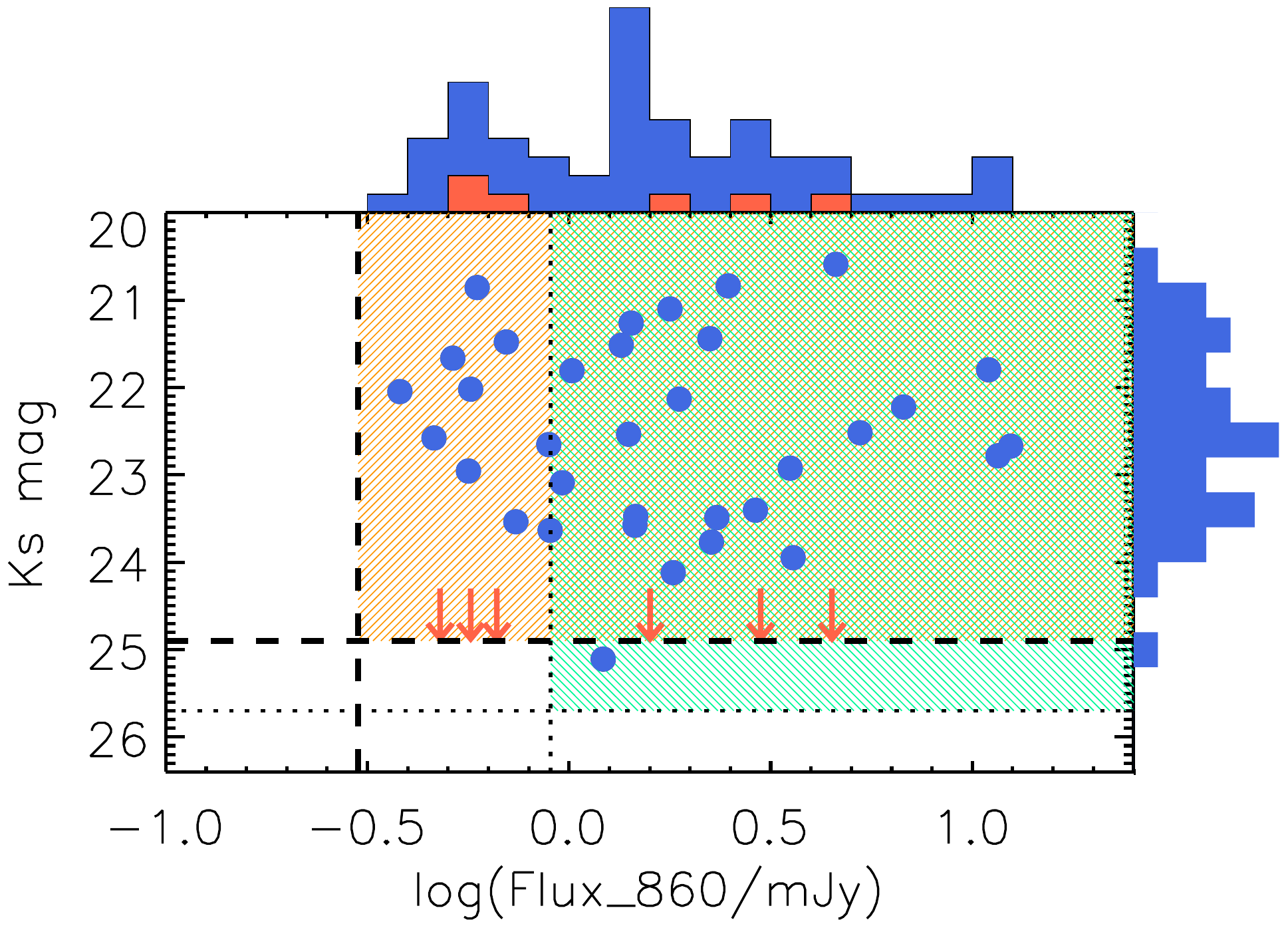}
      \caption {Distribution of observed Ks-band magnitude and 860-$\mu$m ALMA fluxes of ALPINE serendipitous continuum sources with a Ks counterpart. 
      The horizontal dashed line shows the UltraVISTA 
      Ks-band 5$\sigma$ limit of 24.9, while the vertical one shows the minimum 860-$\mu$m (5$\sigma$) flux density reached by the ALMA sample (the orange coloured area
      shows the region covered by our data). For comparison, the dotted lines (green filled area) indicate the Ks magnitude and 870-$\mu$m flux limits of the AS2UDS sample 
      of SMGs reported by \citet{dudzeviciute20}. 
       The downward-pointing arrows shown at the Ks-band limit are the six sources detected only at mid-IR wavelengths. The histograms show the Ks-band magnitude distribution on the right axis, and the 860-$\mu$m flux density 
       distribution is on the top axis of the plot (orange histogram showing the fluxes of the mid-IR detected sources). 
                     }
         \label{Fig:flux_Ks}
   \end{figure}
%----------------------------------------------------------------- 

In Figures~\ref{Fig:optID}, \ref{Fig:optID2}, \ref{Fig:optID3}, and \ref{Fig:optID4} we show some examples of different cases resulting from the identification process of the ALPINE continuum non-target sources: 
from $top$ $left$ to $bottom$ $right,$ we plot the ALMA $>$3$\sigma$ contours superimposed to the ALMA, HST/ACS-i (5$\sigma$$=$27.5 mag in COSMOS; \citealt{koekemoer07}), Subaru (IA484(5$\sigma$)=25.9 mag in COSMOS; \citealt{taniguchi15}), UltraVISTA (Ks(5$\sigma$)$=$24.9 mag in COSMOS; \citealt{mccracken12}, Moneti et al. 2019), IRAC (S$_{4.5}$(5$\sigma$)$\simeq$1.67 $\mu$Jy in COSMOS; \citealt{capak12, steinhardt14}), MIPS (S$_{24}$(5$\sigma$)$=$71 $\mu$Jy in COSMOS; \citealt{lefloch09}), and radio VLA-1.4GHz images 
(5$\sigma$$=$10.5 $\mu$Jy in COSMOS; \citealt{schinnerer10}). 
Figure~\ref{Fig:optID} shows an object with multi-wavelength counterparts in all bands and photo-$z$ from \citet{laigle16}. Figure~\ref{Fig:optID2} shows an object with near-IR to sub-mm identification and photo-$z$. 
Figure~\ref{Fig:optID3} shows an optically$+$near-IR dark galaxy detected only in the SPLASH/IRAC-3.6/4.5~$\mu$m images. Figure~\ref{Fig:optID4} shows an unidentified source.

In Figure \ref{Fig:flux_Ks}, we show the Ks magnitude versus 860-$\mu$m flux for the ALPINE sources with a Ks-band counterpart; the arrows show the locus of the six sources identified only in the mid-IR (plotted
at the 5$\sigma$ limit of the UltraVISTA survey in COSMOS, i.e. Ks$=$24.9). The ALMA flux and Ks
magnitude distribution of our sample are shown on the top and right axes, respectively. We show the ALMA flux and Ks-magnitude limits as vertical and horizontal dashed lines (limiting an orange 
coloured area), and, for comparison, those of the
AS2UDS survey of SMGs (\citet{dudzeviciute20}) as dotted lines (surrounding a green coloured area): while the 860-$\mu$m flux of our sample is significantly fainter than that of 
       the AS2UDS survey, our Ks-magnitude limit is shallower. Our HST$+$near-IR dark sources (shown in orange) span about the whole range in flux, but - by definition - are all below the Ks-band limit (downward-pointing arrows). 
       Only half of them are below the AS2UDS flux limit, the other three could also have been detected in that survey.
      
%----------------------------------------------------------------- 
%----------------------------------------------------------------- 
\subsection{Spectral energy distributions and source properties}
\label{sec:SEDs}
Using all the available photometric data in COSMOS and ECDFS, we constructed the SEDs of all the ALPINE non-target sources with at least one photometric detection in addition to the ALMA one.
%The photometry in the ECDFS is taken directly from the 3D-HST catalogue, providing data from both HST and
%ground-based instruments, already corrected for any biases (e.g., Galactic extinction, PSF size, etc.). 
In order to also obtain the complete mid- and far-IR coverage for our sources, the ALMA
sample was cross-matched with the {\em Spitzer} and {\em Herschel} catalogues in both the ECDFS and COSMOS fields
(i.e. the PACS Extragalactic Probe Survey, PEP, \citealt{lutz11}, the {\em Herschel}-GOODS, H-GOODS, \citealt{elbaz11},
 the Herschel Multi-tiered Extragalactic Survey, HerMES, \citealt{oliver12}, the super-deblended catalogue by \citealt{jin18}).
%The photometry for most of the COSMOS sources is taken from the COSMOS2015 catalogue by \citet{laigle16}, containing also photometric redshifts.
%We also searched for counterparts in the {\em Spitzer} and {\em Herschel} catalogues 
%in COSMOS (i.e., PACS from PEP, \citealt{lutz11}, and SPIRE from HerMES, \citealt{oliver12}).
%Of the sources with no counterparts in COSMOS2015, for 1 we found a match in the 3D-HST, for 3 we obtained the photometry
%from the super-deblended catalogue by \citet{jin18}, while for 4 we obtained the IRAC fluxes from the new SPLASH catalogues by Giulietti et al. (in preparation).
In the COSMOS15 and super-deblended catalogues, the {\em Herschel} fluxes are already reported: we chose the values from the super-deblended catalogue, when available.
No additional {\em Herschel} matches for sources not identified in these two catalogues have been found.
In H-GOODS, the {\em Herschel} fluxes were obtained from IRAC priors, thus source blending should not be an issue. 

For two sources for which a faint counterpart (below the catalogue threshold) is detected only in the IRAC maps, we obtained a magnitude
measurement by performing aperture photometry directly on the images. We remind the reader that the depth of the IRAC/SPLASH observations in COSMOS is 
1.67 $\mu$Jy (5$\sigma$) at 4.5 $\mu$m. Thus, for two sources we obtained IRAC flux densities at 3.6 and 4.5 $\mu$m (S$_{4.5}$$\simeq$1.25 and 1.61 $\mu$Jy, corresponding to $\sim$3.7 and 4.8$\sigma$). 
%while for 1 we derived only a MIPS 24-$\mu$m flux {\bf (45 $\mu$Jy, corresponding to $\sim$3.2$\sigma$)}. 
For six sources (two with just a line identification and no photometric counterparts and four with no counterpart at all -- two of which are likely spurious detections), we could not construct any SEDs. 
%----------------------------------------------------------------- 
%----------------------------------------------------------------- 
\subsubsection{SED fitting}
\label{sec:SEDfit}
We made use of all the available multi-wavelength information (either detections or upper limits) to fit the
SEDs of our sources using the \textsc{Le Phare} software (i.e., {\citealt{arnouts02};
\citealt{ilbert06}), which performs an $\chi^2$ fit to the data by considering different templates. 
We considered the semi-empirical template library of \citet{polletta07}, representative of different classes of IR galaxies
and AGN, to which we added some templates modified in their far-IR part to better reproduce the observed {\em Herschel} data (see
\citealt{gruppioni10,gruppioni13}), and three starburst templates from \citet{rieke09}. 
The final set of templates (32 in total) included SEDs of different types of galaxies, from ellipticals to starbursts, and AGN and composite 
ultra-luminous infrared galaxies (ULIRGs; containing both AGN and star-forming galaxies) in the rest-frame wavelength interval, 0.1--1000 $\mu$m.
We allowed the code to apply different extinction values (E(B--V) from 0.0 to 5) and extinction curves to the templates in order to improve the fit. 
This increased the real number of possible templates.
%----------------------------------------------------------------- 
   \begin{figure*}
   \centering
   \includegraphics[width=17cm]{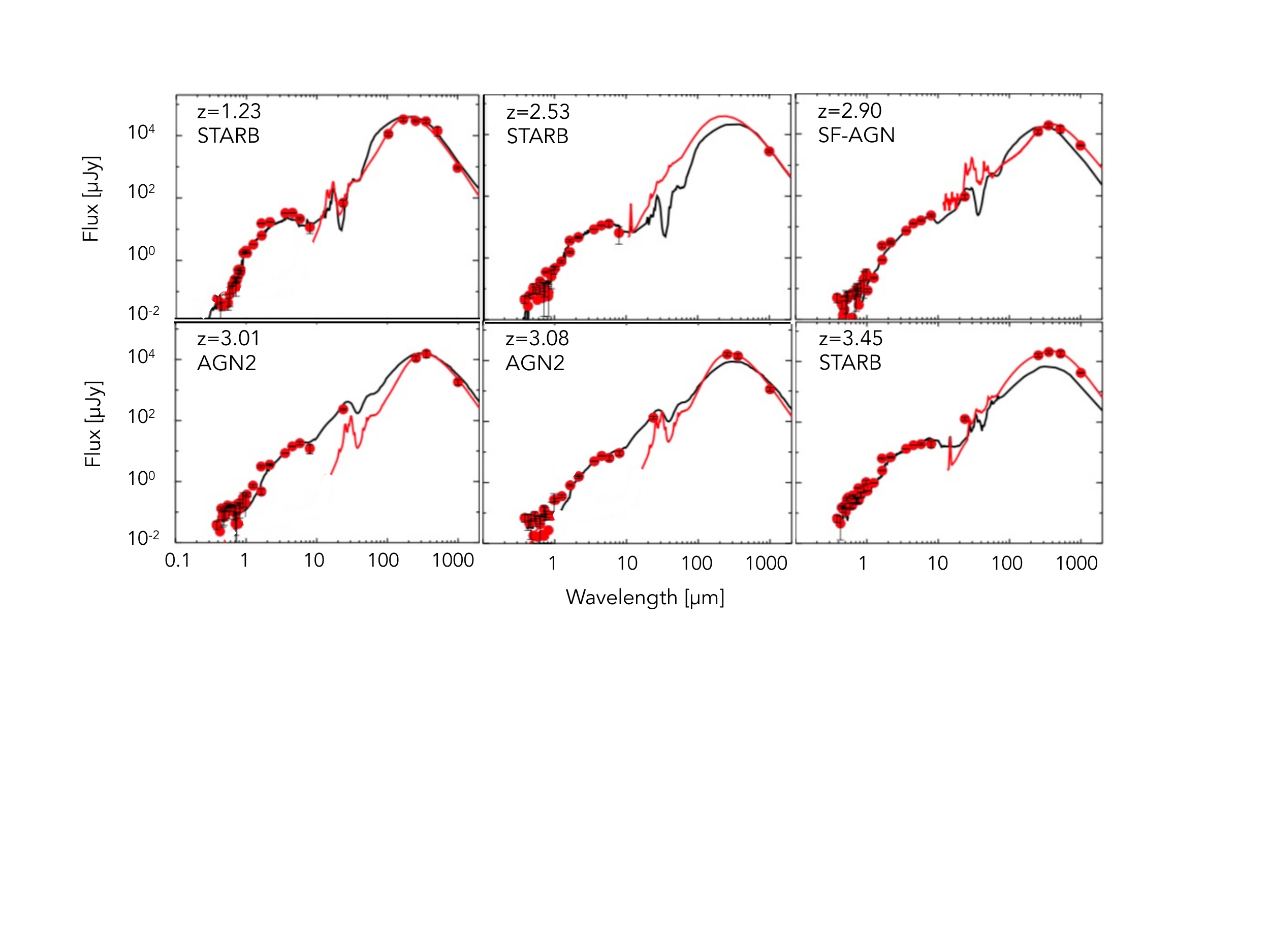}
      \caption{Example of observed SEDs of ALPINE continuum non-target sources with an identification and a photo-z in the available catalogues. 
      The observed SEDs have been fitted with \textsc{Le Phare} by fixing the redshift at the catalogue value: the best fit template to all the data is shown 
      in black, while the template best reproducing the IR data (i.e. from 8 to 1000 $\mu$m rest frame, used to derive $L_{\rm IR}$) is shown in red.
              }
         \label{fig:SEDs}
   \end{figure*}
%-----------------------------------------------------------------
When performing the fit, the redshifts were fixed to the spectroscopic or photometric values from the literature, or from [C~II] line detection, when available. 
In most cases, we found good consistency between the photo-$z$ from the literature and the best fit SED obtained with our SED fitting by fixing the redshift at that value.
For the six sources with only a mid-IR counterpart, we attempted a photo-$z$ estimate with \textsc{Le Phare}, obtaining values of $z_{\rm phot}^{\rm IR}$
in the 2.2--6 range (with an average value $\overline{z}_{\rm dark}$$=$3.7; see Section~\ref{sec:optdark}).
In order to obtain a better determination of the total IR luminosity, we simultaneously fitted only the rest frame's 8-to-1000 $\mu$m range with additional far-IR template libraries included in \textsc{Le Phare}
(e.g. \citealt{chary01,dale02,lagache04,rieke09,siebenmorgen07}). We thus 
best fitted the far-IR bump rather than constraining the whole SED from UV to mm (where optical/near-IR data always dominate the $\chi^2$, because of their smaller errors than those affecting the longer wavelength bands).

For most of the continuum non-target ALPINE galaxies we could obtain a good fit to all the data points and a SED estimate: the majority (75\%) are best reproduced by star-forming galaxy templates
(though 55\% of them are composite, i.e. star-forming galaxies containing an obscured or low-luminosity AGN), while the remaining 25\% are fitted by type 1 or 2 AGN templates.
We checked for signal by stacking on the X-ray images (Chandra) at the positions of the AGN and non-AGN samples, but measured no significant signal for either of the samples. Although,
for the AGN-SED sources, a 1.5$\sigma$ positive signal was detected, against a negative signal for the non-AGN SEDs.
We stress that for the six sources detected only in the mid-IR (i.e. IRAC bands), the SED type and redshift are very uncertain, and the relative results have to be taken only as indications.

In Figure~\ref{fig:SEDs}, we show some examples of the observed SEDs and their best fitting templates obtained from our analysis.
The redshift distribution of the whole sample, including the spectroscopic and photometric redshifts from the literature, those from [C~II] detection,
and those obtained with \textsc{Le Phare} for the sources not in the COSMOS2015, super-deblended, or 3D-HST catalogues, is shown in Figure~\ref{FigZdistr}.
The five redshifts from [C~II] are in a different colour, since we treated those sources separately in the LF analysis because (being at the same
redshifts of the ALPINE targets at the centre of the ALMA pointing) they might be part of an overdensity, or in any case associated to the target. 
Indeed, at $z$$\simeq$4.57, a massive proto-cluster of galaxies located in the COSMOS field has been identified by \citet{lemaux18}, therefore some of
our [C~II] emitters might be part of it.
Considering them as blindly detected sources might bias the LF calculation (see \citealt{loiacono20}). 
These possible effects are discussed in Section~\ref{sec:totLF}.
%----------------------------------------------------------------- 
   \begin{figure}
   \centering
   \includegraphics[width=8.8cm]{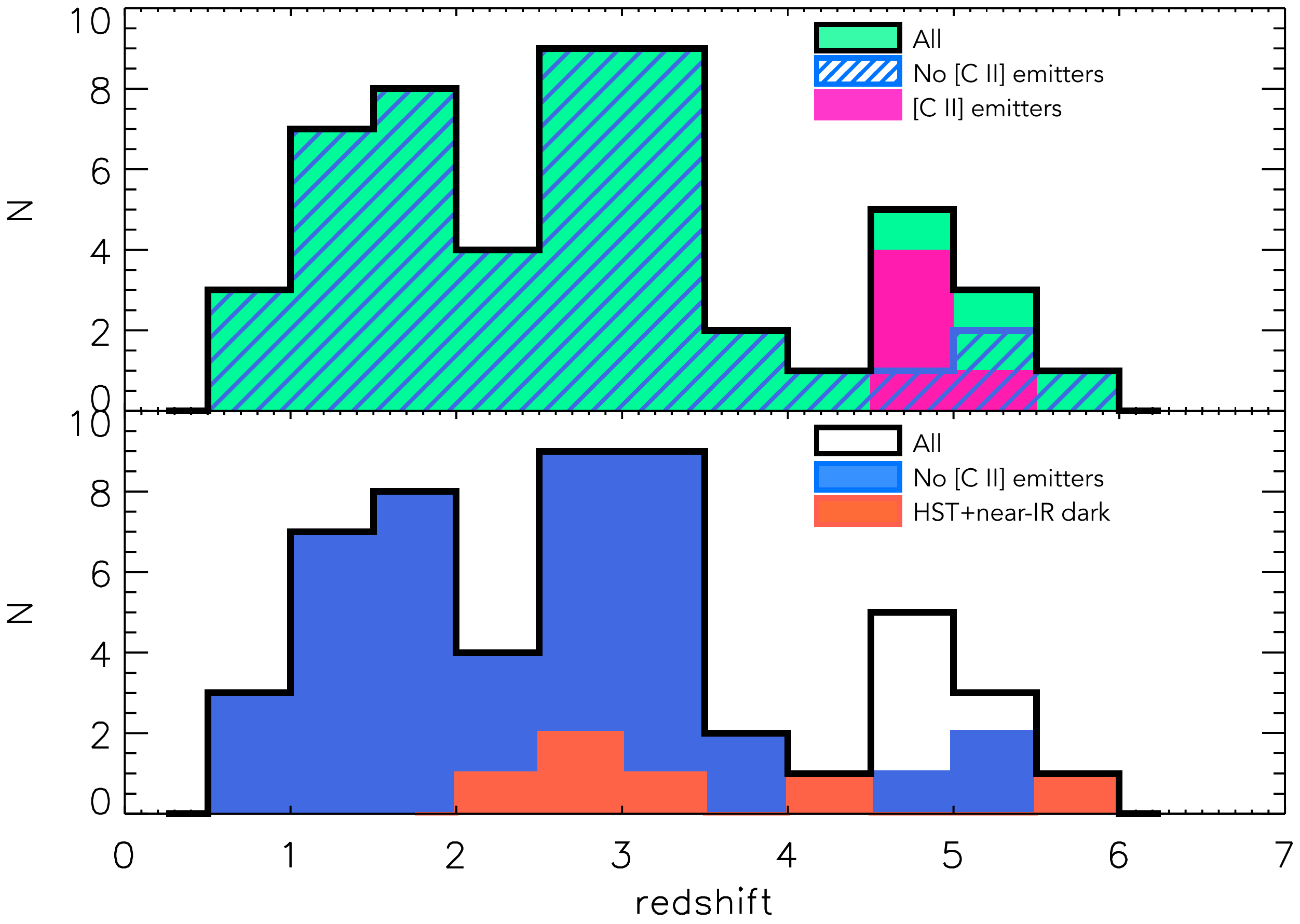}
      \caption{Redshift distribution of ALPINE non-target sources detected in continuum with an identification (green-filled black histogram in the $top$ panel, empty in the $bottom$ panel). 
      In the $top$ panel, the deep-pink histogram shows the five sources detected in [C~II] at the same redshift of the ALPINE central targets, while the blue-dashed histogram shows the redshifts of the 47 sources 
      considered for the unbiased LF calculation (i.e. excluding the five [C~II] emitters). In the $bottom$ panel, the latter distribution is shown in blue, 
      while the best fit photometric redshifts of the six HST$+$near-IR dark galaxies are shown as a red and orange-filled histogram.
              }
         \label{FigZdistr}
   \end{figure}
%-----------------------------------------------------------------
%-----------------------------------------------------------------
   \begin{figure*}
   \centering
   \includegraphics[width=18.5cm]{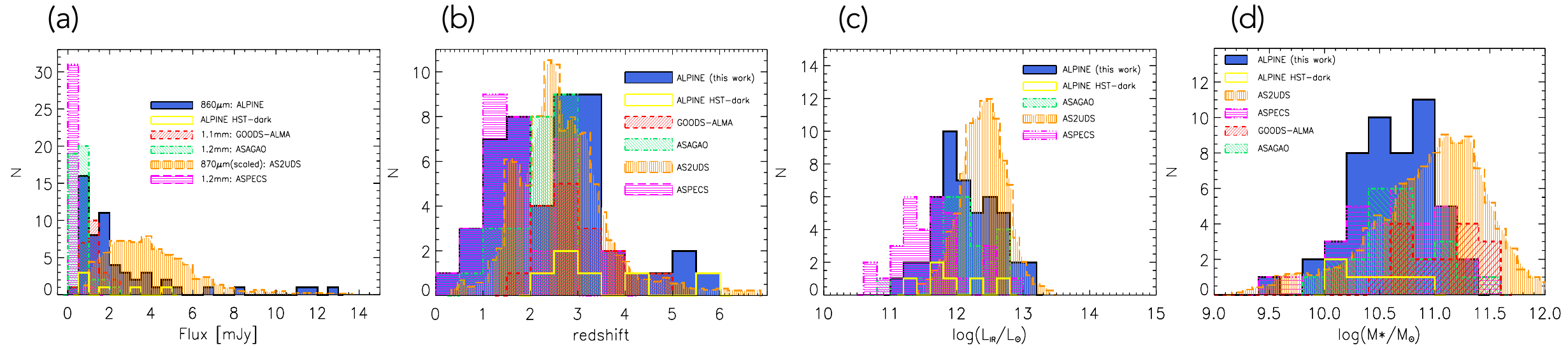}
      \caption{Distribution of 860-$\mu$m flux density of all 56 ALMA continuum serendipitous detections (a), redshift (b), total IR luminosity (c), and stellar mass distribution (d) of the 47 sources 
with measured redshift that were not associated with the ALPINE targets (i.e. blue histogram in Figure~\ref{FigZdistr}). The four distributions are compared with previous results from the literature, either from
blind ALMA surveys like GOODS-ALMA, ASAGAO, and ASPECS (\citealt{franco18, hatsukade18, aravena20}, respectively), or from ALMA surveys of pre-selected SMGs such as the AS2UDS (i.e. \citealt{dudzeviciute20};
the AS2UDS distributions have been rescaled, i.e. divided by 100, for comparison purposes). The ASAGAO masses are from Table 1 of \citet{yamaguchi20}: the plotted values are the ZFOURGE ones (those derived with \textsc{MAGPHYS} are significantly higher). 
The different colours and shadings of the histograms associated with each survey are shown in the legend. The yellow histograms show the distribution of the six HST$+$near-IR dark galaxies. 
We note that the flux densities reported in panel $(a)$ from different surveys are at different wavelengths, therefore not directly comparable, meaning the 860/870-$\mu$m fluxes would correspond to about a factor of $\sim$2 fainter 
values if translated to 1.1/1.2 mm.}
         \label{Fig:flux_z_lir}
   \end{figure*}
%-----------------------------------------------------------------
%----------------------------------------------------------------- 
\subsubsection{Redshifts}
\label{sec:zdist}
In the paper by \citet{bethermin20}, the redshift distribution is presented and discussed only for the `secure' identifications, that is, the 38 sources with a counterpart in the catalogues.
In this work, we used all the redshifts, including the more uncertain ones, considering a total of 52 out of 56 sources. 
The redshift distribution obtained for the whole sample of 52 sources is shown in Figure~\ref{FigZdistr} (green-filled and empty black histogram in the $top$ and $bottom$ panels, respectively).
We note that the total redshift distribution has a broad peak in the $z$$\simeq$1.5--3.5 range (with a low-significance dip at $z$$\sim$2), 
followed by a secondary peak at $z$$\sim$5. %and a tail up to $z$$\simeq$6. 
The secondary peak at $z$$\simeq$5 is partially %mostly 
due to the sources 'associated' with the ALPINE targets (i.e. with a line in the same ALMA side band; deep-pink histogram in the $top$ panel), although it is also occupied by %the higher redshift tail is made by 
sources apparently not related to the targets. 

The median redshift of the total distribution is $z_{med}$$\simeq$2.84$\pm$0.18 ($z_{med}$$\simeq$2.53$\pm$0.17 if we exclude the sources at the same $z$ of the targets; blue-dashed 
and blue-filled distribution in the upper and lower panel, respectively), which is similar to that found by \citet{franco18} in a 2--3$\times$ larger (69 arcmin$^2$) and shallower (to 0.7 mJy) ALMA survey at 1.1 mm in GOODS-S ($z_{med}$$\simeq$2.9), although 
the number of blindly detected objects in our ALPINE pointings is larger (56 compared to 20). 
The size of our continuum survey is similar to that of the  ALMA twenty-six arcmin$^{2}$ Survey of GOODS-S at One-millimeter (ASAGAO; \citealt{hatsukade18}), which is 26 arcmin$^2$, 
although our number of detections is more than double (i.e. we detect 56 sources
above 5$\sigma$ compared to 25 in ASAGAO). However, we must note that our sources are selected in two different side bands, 
and the 1.1 mm one goes about a factor of 1.5--2 deeper than the ASAGAO survey at the same wavelength. 

In panel $(b)$ of Figure~\ref{Fig:flux_z_lir}, the redshift distribution of the 47 identified sources of our sample (excluding the [C~II] emitters) is compared to those from other ALMA surveys, 
such as GOODS-ALMA, ASAGAO,  the ALMA Spectroscopic Survey in the Hubble Ultra Deep Field (ASPECS; \citealt{aravena16, aravena20}) and AS2UDS. While the ASPECS distribution (magenta) peaks at lower redshift, 
and the GOODS-ALMA (red) at slightly higher redshift than ours (blue filled), we notice that the ALPINE continuum survey peaks at redshifts similar to those of the ASAGAO (green dashed) and AS2UDS (orange dashed) surveys. 
However, the secondary peak at $z$$\sim$5 of our distribution does not seem to be present in the other two.
We refer the reader to \citet{bethermin20} for a more detailed discussion about the redshift distribution of the ALPINE continuum non-target sources and the comparison with other ALMA survey works.
%-----------------------------------------------------------------

\subsubsection{Total IR luminosity}
\label{sec:lir_z}
We derived the total IR luminosities ($L_{\rm IR}$$=$$L$[8--1000\,$\mu$m]) for all the ALPINE sources with at least one detection in addition to the ALMA one,
that is, for 50 galaxies (47$+$3 [C~II] emitters). 
In order to obtain the total IR luminosities, we integrate the best fit SED of each source over the range 8$\leq$$\lambda_{\rm rest}$$\leq$1000\,$\mu$m. This integration 
for most of our sources has been performed on well constrained SEDs covered by data in several
bands (see Figure~\ref{fig:SEDs}), while for few sources an extrapolation of the SED with no data constraining the far-IR peak 
was required (thus reflecting in large uncertainties in $L_{\rm IR}$). 
In Figure~\ref{Fig:flux_z_lir}, we show the distribution of the 860-$\mu$m flux density of all the 56 ALMA continuum serendipitous detections (panel $(a)$), and the redshift (panel $(b)$), $L_{\rm IR}$ (panel $(c)$) and stellar mass (panel $(d)$) distributions 
of the 47 sources with spectroscopic or photometric redshift measurements not associated with the ALPINE targets (i.e. blue histogram in Figure~\ref{FigZdistr}), compared to other samples from the literature.
The total IR luminosity distribution of our sources (blue) peaks at $L_{\rm IR}$$\simeq$10$^{12}$ L$_{\odot}$, similarly to ASAGAO (green; \citealt{hatsukade18}), although we cover a larger range, from $\sim$10$^{11}$ to 10$^{13}$ L$_{\odot}$. 
The ASPECS survey (\citealt{aravena20}) peaks at significantly fainter luminosities ($L_{\rm IR}$$\simeq$3$\times$10$^{11}$ L$_{\odot}$), while the AS2UDS (\citealt{dudzeviciute20}) at brighter ones$(L_{\rm IR}$$\simeq$4$\times$10$^{12}$ L$_{\odot}$).
This is a direct consequence of the combination of the flux (panel $(a)$) and redshift (panel $(b)$) histograms. The AS2UDS survey, though covering the same redshift range of ALPINE, extends to and peaks at much larger fluxes, while the ASPECS
survey covers much fainter fluxes and lower redshifts, resulting in fainter luminosities. Also, ASAGAO seems to sample fainter fluxes than ALPINE, but similar redshifts (though it misses a low-$z$ peak present in our distribution), while their $L_{\rm IR}$ are 
similar to ours (with a narrower distribution, but similar peak). The similar luminosity distribution derived from similar redshifts but fainter fluxes might appear puzzling: however, we must note that the flux densities reported in panel $(a)$ of
Figure~\ref{Fig:flux_z_lir} are at different wavelengths, as made clear in the legend, and are therefore not directly comparable. This means that the 860/870-$\mu$m fluxes would correspond to about a factor of $\sim$2--3 fainter 
values if translated to 1.1/1.2 mm. Indeed, our 1-mm channel (here, all the ALPINE flux densities are converted to 860-$\mu$m for simplicity) reaches fluxes 1.5--2$\times$ fainter than ASAGAO. This could likely reconcile our results with the ASAGAO ones.
 
%----------------------------------------------------------------- 
\subsubsection{Stellar mass}
\label{sec:mass}
We used the \textsc{Le Phare} code and the \citet{bruzual03} libraries to estimate the stellar masses of our sources.
We stress that the stellar masses derived for the HST and near-IR dark galaxies are extremely uncertain, given the few photometric points available
(although we made use of all the 3$\sigma$ upper limits to constrain the masses). 
Therefore, we can only take the results as an indication.
In panel $(d)$ of Figure~\ref{Fig:flux_z_lir}, we show the mass distribution of our sources (blue) compared to those from GOODS-ALMA (red), ASAGAO (green), ASPECS (magenta) and AS2UDS (orange).
We find that our galaxies are massive (our mass distribution extends up to masses as high as 2.5$\times$10$^{11}$ M$_\odot$; see also Figure~\ref{fig:z_mass}), but slightly less extreme than those detected by \citet{franco18} and
by \citet{dudzeviciute20}. In fact, the median stellar mass of our distribution is M$^*$$=$(4.1$\pm$0.7)$\times$10$^{10}$ M$_\odot$, while for GOODS-ALMA is 1.1$\times$10$^{11}$ M$_\odot$ and 
for AS2UDS (1.26$\pm$0.05)$\times$10$^{11}$ M$_{\odot}$. 
The stellar mass distribution of the ASPECS sample is similar to ours, both in peak position and mass coverage.
The ASAGAO masses have been derived by \citet{yamaguchi20} with \textsc{MAGPHYS} (\citealt{dacunha08}) and also by
\citet{straatman16} in the {\tt FourStar} galaxy evolution survey (ZFOURGE) using \textsc{FAST} (\citealt{kriek09}). In the figure we plot the latter ones, which are well consistent with 
ours (while the \textsc{MAGPHYS} masses are significantly higher by $\gtrsim$0.2–0.5 dex, as also noted by \citealt{yamaguchi20}).   
%----------------------------------------------------------------- 
%----------------------------------------------------------------- 
   \begin{figure}
   \centering
   \includegraphics[width=9cm,height=5cm]{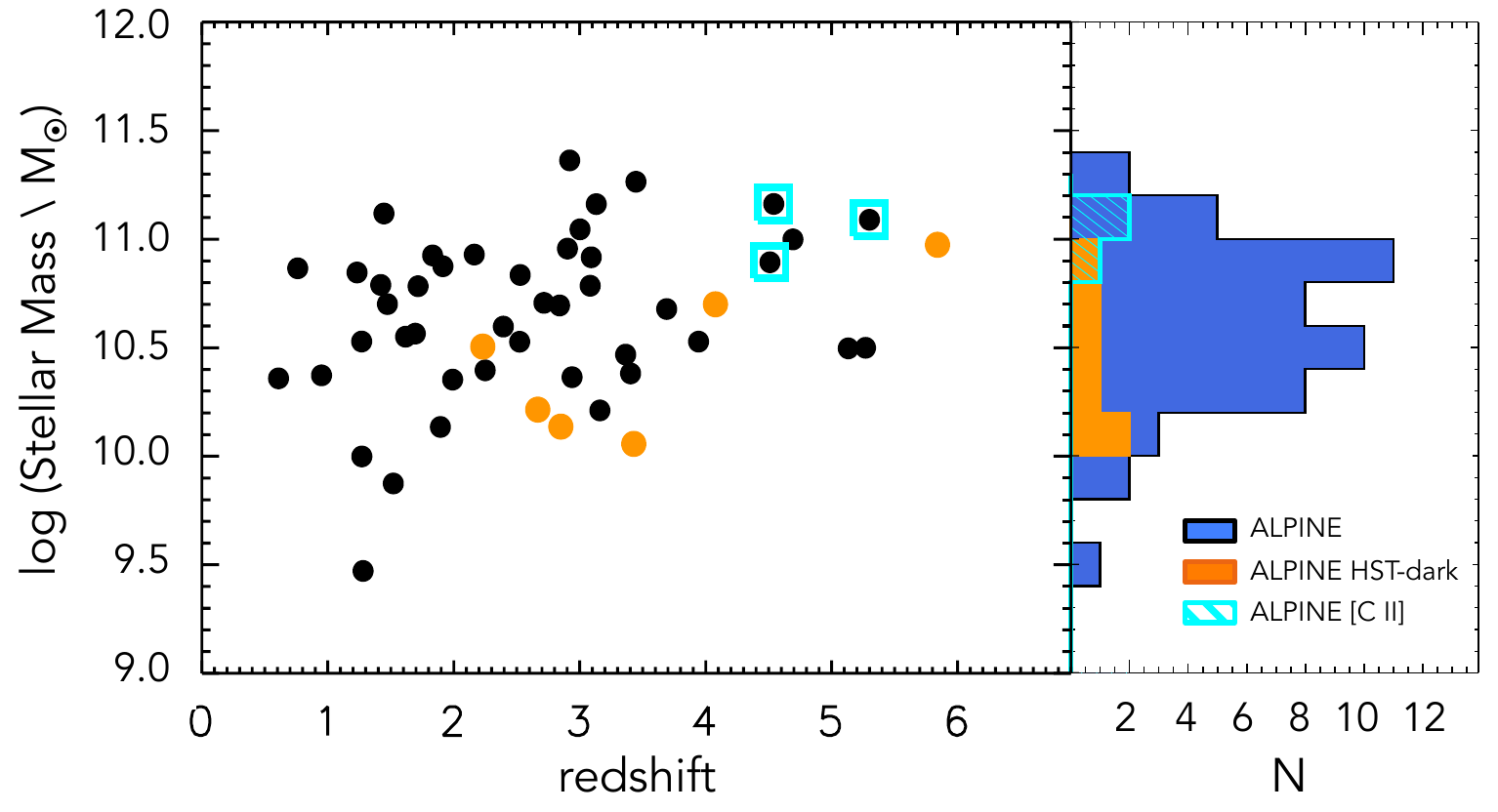}
      \caption{Stellar mass versus redshift ($left$) and stellar mass histogram ($right$) for ALPINE continuum non-target detections. The black filled circles and blue filled histogram show
      the distribution of the whole sample, while the orange circles and histogram show the locus occupied by the HST$+$near-IR dark sources. The cyan open squares and dashed-histogram 
      show, for comparison, the locus of the sources identified with [C~II] emitters at the same redshift of the ALPINE targets (we note that the two without photometric counterparts are missing, since for them a mass estimate was impossible).}
         \label{fig:z_mass}
   \end{figure}
%-----------------------------------------------------------------
\subsection{Optically and near-IR dark galaxies}
\label{sec:optdark}
%----------------------------------------------------------------- 
   \begin{figure}
   \centering
   \includegraphics[width=9cm,height=10cm]{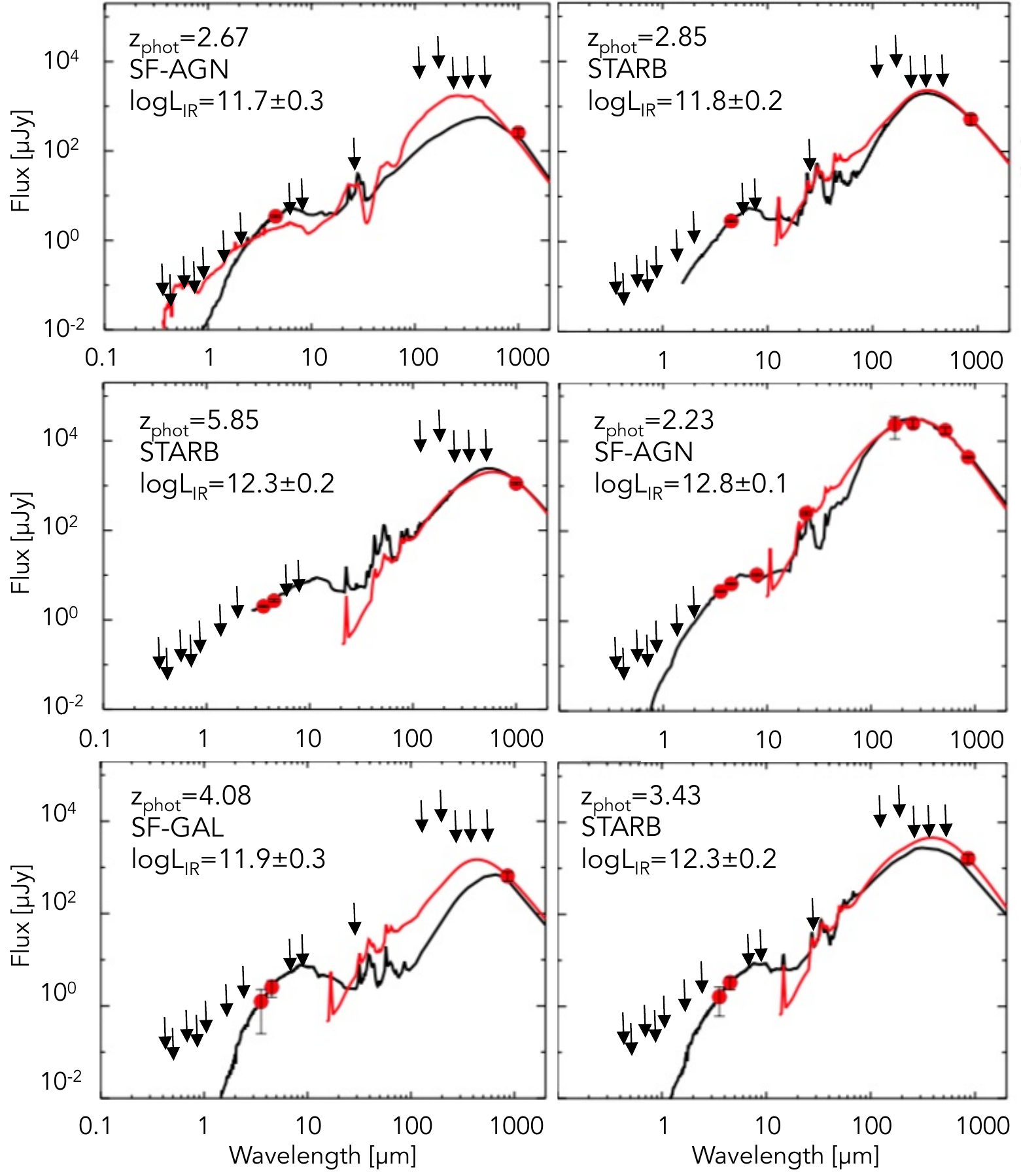}
      \caption{Observed SEDs for HST$+$near-IR-dark galaxies, with their tentative best fitting templates (black for the broad-band SED and red for the far-IR SED, 
      as in Figure \ref{fig:SEDs}) and photometric redshifts found with \textsc{Le Phare}. For all but one source, the fits are based only on the IRAC and ALMA data points, combined with the 3$\sigma$
      UV, optical, near- and far-IR 3$\sigma$ upper limits. }
         \label{fig:SED_dark}
   \end{figure}
%-----------------------------------------------------------------
As mentioned in the previous section, of the 56 galaxies detected in our main catalogue, 12 (21\%) do not present any obvious HST or near-IR (UltraVISTA,
to $Ks$$\simeq$24.9; see \citealt{mccracken12}, DR4: Moneti et al. 2019) counterparts. 
Six of these sources have been identified in the IRAC 3.6 or 4.5 $\mu$m bands (one also in the MIPS 24-$\mu$m and {\em Herschel} bands), while six have no photometric counterpart at all.
Two of these unidentified sources have been detected as line (likely [C~II]) emitters by \citet{loiacono20}, while four remain unidentified (compatible with the number of expected 
spurious sources based on simulations, though two are at significantly high SNR; see \citealt{bethermin20}). 
If we exclude the four sources without identification, we end up with a $\simeq$14\% fraction of HST$+$near-IR dark galaxies among the ALPINE non-target continuum detections (six with a mid-IR counterpart $+$ two [C~II] emitters).
The observed SEDs of the six dark galaxies with an IRAC (or far-IR) counterpart, and their best fit template found by
\textsc{Le Phare}, are shown in Figure~\ref{fig:SED_dark}. 
%The photometric redshifts derived by considering only the mid-IR (in one case also far-IR) and ALMA data points and the upper limits in the other bands, 
Their photometric redshifts are in the 2.23--5.85 range, with an average value of $\overline{z}_{\rm dark}$$\simeq$3.5$\pm$0.5.
We stress again that the estimated redshifts for five of these sources are extremely uncertain (while for the reminder the photo-$z$ is better constrained by the available IRAC, MIPS and {\em Herschel} data); 
the width of the probability distribution function (PDF) can be as large as $\sim$1--1.5. Moreover, with few photometric data, the best fitting solutions can degenerate in the photo-$z$/$A_V$ space (i.e. \citealt{caputi12}).
However, in our case the ALMA detection and the absence of optical and near-IR counterparts come to our aid, as they ruled out the low-$z$ solutions and better constrained the photometric redshift.
%----------------------------------------------------------------- 
   \begin{figure*}
   \centering
   \includegraphics[height=12cm]{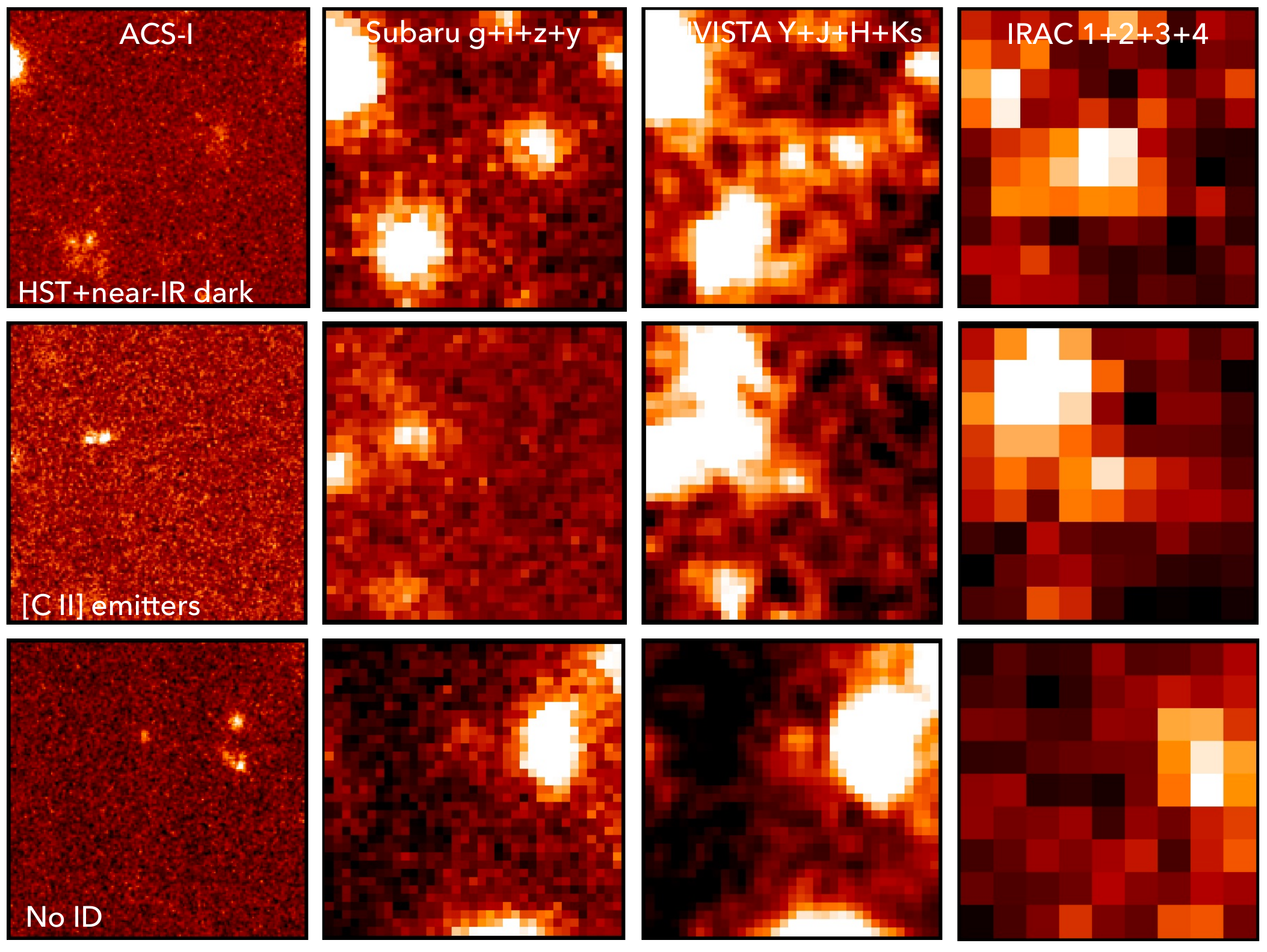}
      \caption{Stacked images (10 arcmin size) resulting from co-addition of ACS-I, Subaru $g+i+z+y$, Ultra-VISTA $Y+J+H+Ks,$ and IRAC $ch1+ch2+ch3+ch4$ bands (from $left$ to $right,$ respectively) at the positions of the 
      six HST$+$near-IR dark galaxies ($top$), of the two [C~II] emitters without photometric counterparts ($middle$), and of the four unidentified sources ($bottom$). }
         \label{fig:stack}
   \end{figure*}
%-----------------------------------------------------------------

In order to check whether we can detect and eventually measure an average flux in the optical and near-IR bands for these dark sources, we performed stacking at their positions in the HST-ACS band
and in all four Subaru, Ultra-VISTA, and IRAC bands. We note that in the Subaru, Ultra-VISTA, and IRAC bands, we co-added images at different wavelengths. We thus applied average 
corrections to the fluxes when required (i.e. we multiplied the 2$^"$ aperture photometry value of the IRAC stacked flux by a factor 1.22). In fact, the aim of our stacking analysis was not to measure accurate values, 
but just to validate our conclusions by detecting and estimating an average flux density for the ALPINE galaxies undetected in the optical and near-IR.
In Figure~\ref{fig:stack}, we show the results of our stacking analysis for the six HST$+$near-IR dark galaxies ($top$ row) in the ACS-I, Subaru ($g+i+z+y$), UltraVISTA ($Y+J+H+Ks$), and IRAC ($ch1+ch2+ch3+ch4$) bands 
(from left to right, respectively). We compared to the results obtained for the two [C~II] emitters without any counterparts ($middle$ row) and for the four unidentified sources ($bottom$ row).
A positive signal comes up clearly in the UltraVISTA and IRAC bands for the six HST$+$near-IR dark galaxies, providing an average flux of %(1.25$\pm$0.08) and (2.58$\pm$0.18) $\mu$Jy respectively.
$\sim$(1.21$\pm$0.03) and (1.7$\pm$0.3) $\mu$Jy, respectively. 
%A barely visible signal (at $\sim$2$\sigma$) appears at the centre of the Subaru stacked image, while in 
In the ACS and Subaru images, we detect only the background. 
The images co-added at the positions of the two [C~II] emitters without counterparts show a faint signal only in the UltraVISTA bands, and maybe in the IRAC ones, but nothing in the
ACS and Subaru bands. The four unidentified galaxies do not show any signal in the stacked images, at any wavelengths.
In a future paper (Gruppioni et al. in preparation), we intend to investigate and discuss in more detail the nature and average properties of the ALPINE optical and near-IR dark sources (detected both in continuum and [C~II]).

Previous studies have found faint ALMA galaxies completely missed at 
optical and near-IR wavelengths  (\citealt{franco18, wang19,yamaguchi19}), even in the deepest  surveys in GOODS. 
The fraction of HST-dark sources discovered in the GOODS-ALMA survey by \citet{franco18} is 20\% of their sample, 
which is similar to the fraction found for the ASAGAO survey by \citet{yamaguchi19}, 
and similar to the fraction that we find in our sample if we consider all the sources not detected in the HST and near-IR bands$(\sim$21\%). 
However, if we exclude the four sources without any counterparts, but keep the two sources with [C~II] lines, we find a more realistic percentage of 14\%. 
If we also exclude the two [C~II] emitters, likely associated with the ALPINE targets, 
we find a conservative percentage of 11\% of serendipitous HST$+$near-IR dark galaxies in our sample. 
However, for a fair comparison, we must note that the HST-dark galaxies found by \citet{franco18} are undetected in GOODS-S, whose photometry is deeper 
than in COSMOS. Therefore, some of our HST-galaxies could have been detected in optical or near-IR images as deep as the ones covering the GOODS-S field. 
Indeed, this would further reduce our fraction of HST-dark sources, increasing the difference with the previous results. For a more direct comparison of the results, in Table~\ref{tab:hstdark} we report the size and depth of ALPINE, GOODS-ALMA, and ASAGAO.
%-----------------------------------------------------------------
\begin{table}
\caption{ALMA selected HST-dark sources}             % title of Table
\label{tab:hstdark}      % is used to refer this table in the text
\centering                          % used for centering table
\footnotesize
\begin{tabular}{c c c c c c}        % centered columns (4 columns)
\hline\hline               % inserts double horizontal lines
Survey & Area & S$_{lim}$ (5$\sigma$) & $\lambda$ & Ntot &  \%dark \\ 
      &  (arcmin$^2)$ &   (mJy)  &  ($\mu$m) &  &  (\%) \\
      (1) & (2) & (3)  & (4) & (5)    & (6)      \\  % table heading 
\hline    
   ALPINE &  24.9 &  0.3  &  860  & 56 & 11(21)\tablefootmark{a}  \\     % inserting body of the table
   GOODS-ALMA  & 69 &  0.70 & 1100 & 20 & 20   \\
   ASAGAO & 26 & 0.2 & 1200 & 25 & 20 \\
 
\hline     \hline                             %inserts single line
\end{tabular}
\tablefoot{
\tablefoottext{a}{The value in parentheses is the maximum (unrealistic) percentage obtained by considering all the sources unidentified in the optical and near-IR, including the likely spurious ones. }}
\end{table}

%-----------------------------------------------------------------
While the depth and size of the GOODS-ALMA survey are different to ours (it is about 2.5$\times$ the size and 2--3$\times$ shallower),
the ASAGAO survey is similar to ALPINE, both in size and sensitivity. However, our detections are either at 860 or 1000 $\mu$m, while the
two mentioned surveys in GOODS-S are at 1100-1200 $\mu$m. A similar depth but in two different selection bands, in a range where the galaxy SEDs are steep, 
makes our survey about 2$\times$ deeper than the ASAGAO survey. 
Given all these factors (ALPINE deeper in ALMA, but shallower in the counterpart identification), we would have expected to find a larger fraction of galaxies 
undetected in the HST and/or UltraVISTA bands in ALPINE (COSMOS) than in GOODS-ALMA or ASAGAO. 
However, we must note that, considering the shot noise, the uncertainties in equivalence of detection and matching methodology, the data quality and depth in various bands, 
we cannot take this as a really significant difference.

The stellar masses estimated for the HST$+$near-IR dark galaxies in our sample (shown in orange in Figure~\ref{fig:z_mass} and as a yellow histogram in panel $(d)$ of Figure~\ref{Fig:flux_z_lir}) 
span about an order of magnitude in stellar mass, from 1.1$\times$10$^{10}$ to 9.5$\times$10$^{10}$ M$_\odot$, and are not necessarily the most massive of the sample.
For the purpose of the luminosity function calculation, we considered the HST-dark galaxies, although with large uncertainties in their
redshifts and 8--1000\,$\mu$m integrated luminosities (accounted for in our simulations).
%----------------------------------------------------------------- 
%----------------------------------------------------------------- 
  \begin{figure*}
   \centering
\includegraphics[width=18cm]{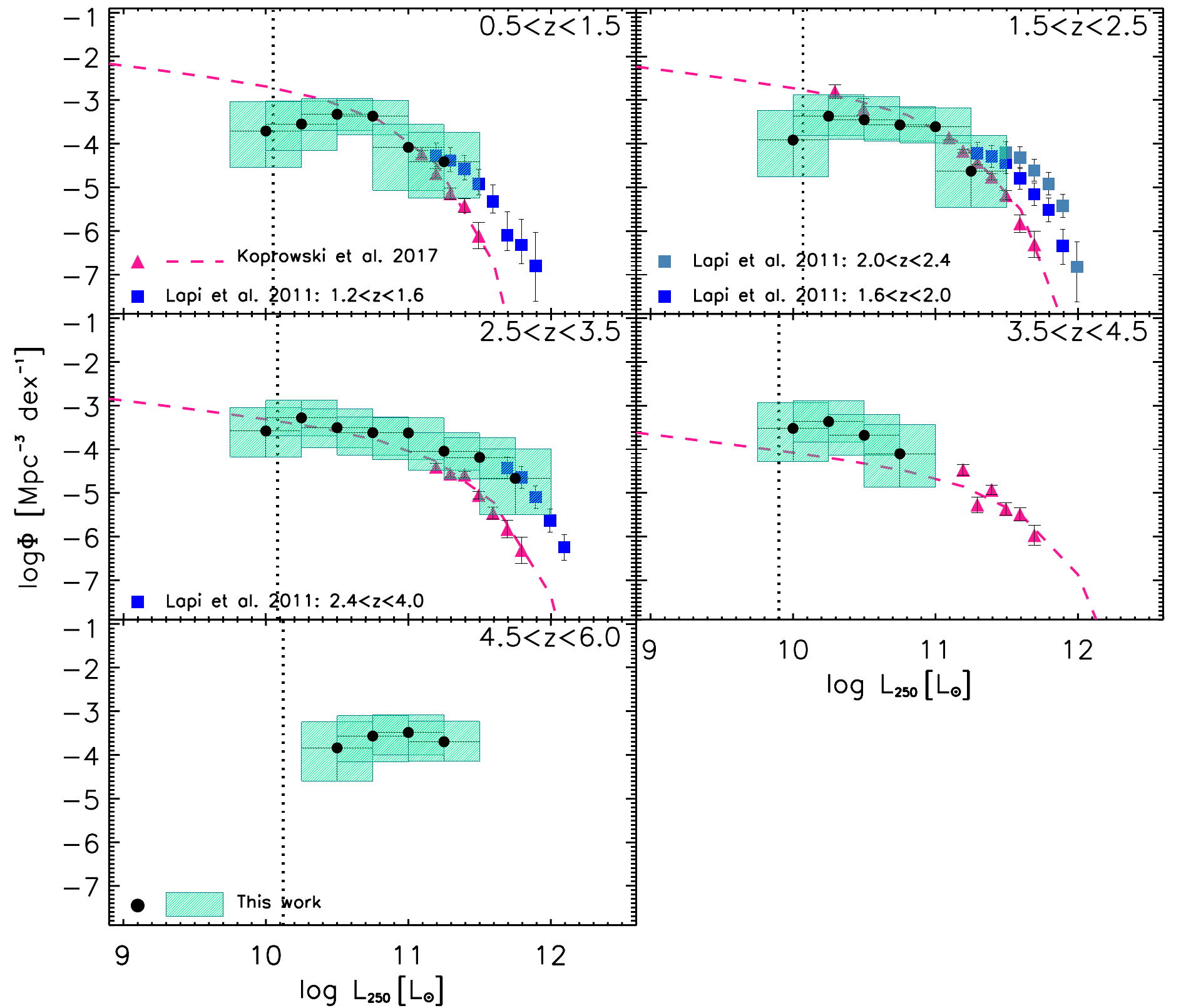}
   \caption{Rest-frame 250\,$\mu$m LF estimated with 1/V$_{\rm max}$ method from ALPINE continuum sample (green boxes and black filled circles). 
The luminosity bins have a width of 0.5 dex in ${L}_{\rm 250 \mu m}$, and step through the luminosity range in stages of 0.25 dex. For this reason, the individual bins are not statistically independent. The error bars in the data points represent the uncertainties obtained from the simulations (as described in Section~\ref{sec:totLF}). 
The deep-pink triangles and dashed curves are the SCUBA-2 250-$\mu$m LFs by \citet{koprowski17}, while the blue filled squares are the {\em Herschel} 
ATLAS 250-$\mu$m LFs by \citet{lapi11}, the latter being in slightly different redshift intervals. The vertical dotted line shows the completeness limit of our continuum survey, estimated as described in the text by considering 
the nominal 860-$\mu$m limiting flux of 0.3 mJy (\citealt{bethermin20}).   }
              \label{fig250LF}
    \end{figure*}
%----------------------------------------------------------------- 
%----------------------------------------------------------------- 
\section{Luminosity function}
\label{sec:lf}
The size and depth of our sample allow us to derive the far-IR LF in more than one redshift bin, spanning from $z$$\simeq$0.5 up to $z$$\sim$6. 
Because of the redshift range covered by our continuum sample, we would need to make significant wavelength extrapolations when computing the rest-frame LFs at any chosen wavelength. 
In order to apply the smallest extrapolations for the majority of our sources, we chose to derive the far-IR LF at the rest-frame wavelength corresponding 
to the median redshift of the sample ($\sim$3): we therefore derive the rest-frame LF at 250\,$\mu$m. 
Given the excellent multi-wavelength coverage of our fields, the SEDs of most of our sources are very well determined from the UV to the sub-mm. 
The extrapolations are therefore well constrained by accurately defined SEDs, even at redshifts lower and higher than the median value. However, there are
few sources for which the photometric redshift is based only on the ALMA and one or two mid-IR fluxes, therefore the redshift itself is very uncertain
and the SED not well sampled. The extrapolation for these sources is thus not very accurate and the luminosity is derived with a high level of indeterminateness (i.e. it may vary by a factor of 2--3). 
We took into account these uncertainties in the error bars associated with the LF values (as discussed in detail in Section~\ref{sec:totLF}).
%----------------------------------------------------------------- 
\subsection{Method}
\label{sec_method}
%----------------------------------------------------------------- 
 %  \begin{figure}
 %  \centering
  % \includegraphics[width=9cm]{eff_area.pdf}
    %  \caption{Effective area of the ALPINE non-target sample as a function of the
%850-$\mu$m flux (after excluding the the central 1-arcsec radius area). The
%blue, green, orange, and red lines are the results obtained for a source size
%of 0, 0.3, 0.6, and 1 arcsec, respectively. The method used to compute
%the surface area is described in Bethermin et al. (2019), from which the figure is drawn.
   %           }
     %    \label{Fig:compl}
  % \end{figure}
%-----------------------------------------------------------------
We derived the LFs using the $1/V_{\rm max}$ method (\citealt{schmidt68}). This method is non-parametric and does not require any 
assumptions on the LF shape, but derives the LF directly from the data. 
In order to derive the monochromatic and total IR LFs, we used all the sources with a spectroscopic or photometric redshift, with the exception of
two sources (SC$\_$1$\_$DEIMOS$\_$COSMOS$\_$787780,
SC$\_$1$\_$vuds$\_$cosmos$\_$5101210235) that were also excluded from the continuum number counts by \citet{bethermin20} because their flux density was found to be boosted by 
an emission line (CO(7-6) at $z$$=$1.28 and [C~II] at $z$$=$4.51, respectively). 
At $z$$\simeq$5, we computed (and compared) two different LFs by either excluding or including the three [C~II] emitters with optical/near-IR identification likely associated with the ALPINE targets. In the case of the former, we used 46 sources (out of 47, one was excluded because it was boosted by a CO line), and for the latter we used 48 (two [C~II] emitters could not be used because they had no counterpart, one was excluded because it was boosted by the line).%, which we spread over all the redshifts.
We divided the sample into five different redshift bins (over the 0.5$\lesssim$$z$$\lesssim$6 range), which we selected to be similarly populated. 
In each redshift bin, we computed the co-moving volume available to each source belonging to that bin, defined as
\begin{equation}
{\rm V_{\rm max,i}=\int_ {\rm z_{min}}^{\rm z_{max}} {\Omega_{\rm eff,i}\frac{\rm dV}{\rm d\Omega dz} dz}=V{\rm(z_{max,i})}-V{\rm(z_{min,i})}}
,\end{equation}
where $z_{\rm max}$ 
is the minimum between the maximum redshift at which a source would still be included 
in the sample -- given the limiting flux of the survey -- and the upper boundary of the considered redshift bin. Analogously, $z_{min}$ is 
the maximum between the minimum redshift above which the source will be detected in the survey and the 
lower boundary of the redshift bin. The quantity $dV /(d\Omega dz)$ is the co-moving volume element per
unit solid angle and unit redshift, while $\rm \Omega_{\rm eff,i}$ is the effective area of the $i$-${\rm th}$ galaxy 
and depends on both the flux density (i.e. the total area covered by the survey, 24.92 arcmin$^2$, at bright fluxes, since only the brightest sources can be 
detected when distant from the centre of the pointing) and the size of each source (e.g. compact sources show better
completeness than extended ones at a given flux density). We note that to calculate the areal coverage of the serendipitous detections, we excluded the 1 arcsec 
central area where the target source was extracted.
The effective area is derived from the completeness Compl(S$_{\rm 850}$,$\theta_{\rm i}$,x$_{\rm i}$, y$_{\rm i}$) at the position (x$_{\rm i}$,y$_{\rm i}$)
of the {\em i}-th source:
\begin{equation}
\Omega_{\rm eff,i}(S_{\rm 850},\theta_{\rm i}) = \sum_{pointings} \int{\int{\rm Compl(S_{850}, \theta_i, x_i, y_i)~d\Omega}}
,\end{equation}
where the sum is over the 118 pointings.
The completenesses were derived through accurate simulations by \citet{bethermin20}.
Their Figure~8 shows the effective area as a function of the 850-$\mu$m flux for different source sizes.

For each luminosity and redshift bin, the LF is given by
\begin{equation}
\centering
\phi(L,z)=\frac{1}{\rm \Delta \log L}\sum_{\rm i}{\frac{1}{\rm V_{max,i}}} \times {\rm incompl}(z)
,\end{equation}
where $\Delta logL$ is the size of the logarithmic luminosity bin, $incompl$($z$) is the correction for redshift incompleteness (i.e. sources without redshift),
and $V_{\rm max,i}$ is the maximum co-moving volume over which the $i$-${\rm th}$ galaxy could be observed given 
its luminosity and redshift (Equation 1).
We adopted $incompl$($z$)$=$1 for $z$$\leq$6, under the assumption that the unidentified sources are all at $z$$>$6 or spurious.
In any case, whether or not we consider the redshift incompleteness (e.g. assuming that the three unidentified sources are at $z$$>$3) will not affect our conclusions.

Uncertainties in the infrared LF values depend on the number of sources in the luminosity bin (i.e. Poissonian error) and on the photometric redshift uncertainties. 
In particular, significant errors on the redshift estimate can shift a low redshift galaxy to higher redshifts and vice versa, thus modifying the number density of sources in a given redshift bin. 
To study the impact of these uncertainties on the inferred IR LF, 
we performed Monte Carlo simulations, as described in Section~\ref{sec:totLF}.
%----------------------------------------------------------------- 
%----------------------------------------------------------------- 
\subsection{The rest-frame 250-$\mu$m luminosity function}
\label{sec_rflf}

Following the method described above, we derived the 250-$\mu$m LF of the ALMA ALPINE sources.
We divided the samples into five redshift bins: 
0.5$<$$z$$\leq$1.5, 1.5$<$$z$$\leq$2.5, 2.5$<$$z$$\leq$3.5, 3.5$<$$z$$\leq$4.5, and 4.5$<$$z$$\leq$6. 
We considered luminosity bins of 0.5 dex, covering the whole luminosity range by overlapping by 0.25 dex. In this way,
the luminosity bins are not statistically independent (they are `alternately' independent), but we can better observe the 'shape' of the
LF and the position of the sources within the bin (e.g. if the bin is uniformly populated or the sources are grouped at the edge of a bin).
To study the possible bias introduced by the sources with spectroscopic redshifts (from [C~II] 158 $\mu$m line emission) very close to that of the
ALPINE targets, we derived two LFs at 4.5$<$$z$$<$6: one by excluding and another by including the two [C~II] emitters with identification from/in the calculation.
The comparison between the two LFs (excluding and including the two sources) is presented and discussed
only in Section~\ref{sec:totLF}, in order to avoid repetitions.

The results of the computation of our rest-frame 250-$\mu$m LFs are reported in Table~\ref{tab:250LF}; the errors were
computed through Monte Carlo simulations in order to study the impact of redshift uncertainties on the LFs. We refer the reader to the next section for a
detailed description of the simulation. Given the area covered by our survey and the number of independent
pointings, the contribution due to cosmic variance  (from \citealt{driver10}) is always negligible with respect to the uncertainties due to
photo-$z$ and luminosity.

Our 250-$\mu$m LFs are shown in Figure~\ref{fig250LF}. The completeness limits, shown as vertical dotted lines, were computed by considering
the nominal 860-$\mu$m limiting flux of our survey, that is, 0.3 mJy (see \citealt{bethermin20}), and the template, among those of the library reproducing our SEDs, which provided the brightest luminosity at the redshift of the bin.
%Note that in the 4.5--6.0 redshift bin we have reported only the LF derived by excluding the 5 
%sources likely associated to the ALPINE targets, leaving the comparison with that obtained by including them and the discussion to the next section. 
\begin{table*}
 \caption{ALPINE rest-frame 250-$\mu$m LF}
\begin{tabular}{cccccccc}
\hline \hline
 log(L$_{250}$/L$_{\odot}$) & \multicolumn{6}{c}{log($\phi$/Mpc$^{-3}$ dex$^{-1}$)   [N$_{\rm obj}$]}\\ \\
                                               & 0.5$<$$z$$<$1.5  & 1.5$<$$z$$<$2.5 & 2.5$<$$z$$<$3.5 & 3.5$<$$z$$<$4.5 & \multicolumn{2}{c}{4.5$<$$z$$<$6.0}   \\ 
                                                &   &  &  &  & No [C~II] emitters & All    \\ 
(1) & (2) & (3)  & (4) & (5)     & (6)    & (7)        \\   \hline 
         \\ \vspace{0.2cm}
%    \em{9.50 -- 10.00} &                                                     &                                                 & \em{$-$2.90$^{+0.41}_{-0.43}$~[1]} &                                                   &                                                   \\ \vspace{0.2cm}   
   \em{9.75 -- 10.25} &    {\em ($-$3.71$^{+0.68}_{-0.83}$~[1])\tablefootmark{a}}  & {\em ($-$3.92$^{+0.68}_{-0.83}$~[1])~} & {\em ($-$3.58$^{+0.54}_{-0.59}$~[2])~} &  {\em $-$3.52$^{+0.59}_{-0.76}$~[1]}  &                                &     \\ \vspace{0.2cm}
   \bf{10.00 -- 10.50\tablefootmark{b}}   &  {\bf $-$3.51$^{+0.54}_{-0.61}$~[2]} &  {\bf $-$3.37$^{+0.49}_{-0.43}$~[4]} &  \bf{$-$3.28$^{+0.40}_{-0.41}$~[5]} &   \bf{$-$3.36$^{+0.48}_{-0.47}$~[2]} &                     &    \\ \vspace{0.2cm}
   \em{10.25 -- 10.75} &  {\em $-$3.33$^{+0.36}_{-0.37}$~[6]} &  {\em $-$3.45$^{+0.53}_{-0.45}$~[4]}  & {\em $-$3.51$^{+0.42}_{-0.45}$~[4]} &  {\em $-$3.68$^{+0.48}_{-0.45}$~[2]} & {\em $-$3.83$^{+0.59}_{-0.76}$~[1]} & {\em $-$3.86$^{+0.59}_{-0.76}$~[1]} \\ \vspace{0.2cm}
 \bf{10.50 -- 11.00} &  {\bf $-$3.37$^{+0.36}_{-0.43}$~[6]} &  {\bf $-$3.57$^{+0.42}_{-0.36}$~[7]}  &  {\bf $-$3.62$^{+0.35}_{-0.51}$~[7]} & {\bf $-$4.10$^{+0.68}_{-0.76}$~[1]} & {\bf $-$3.57$^{+0.47}_{-0.58}$~[2]} & {\bf $-$3.59$^{+0.49}_{-0.52}$~[2]}\\ \vspace{0.2cm}
  \em{10.75 -- 11.25} &  {\em $-$4.08$^{+0.52}_{-0.98}$~[2]} &  {\em $-$3.61$^{+0.43}_{-0.36}$~[7]} &  {\em $-$3.62$^{+0.35}_{-0.62}$~[8]} &                                                            & {\em $-$3.49$^{+0.41}_{-0.51}$~[3]} & {\em $-$3.52$^{+0.46}_{-0.48}$~[3]} \\  \vspace{0.2cm}
   \bf{11.00 -- 11.50} &  {\bf $-$4.41$^{+0.68}_{-0.83}$~[1]} &  {\bf $-$4.63$^{+0.81}_{-0.83}$~[1]} &  {\bf $-$4.04$^{+0.42}_{-0.44}$~[4]} &                                                  &{\bf $-$3.70$^{+0.47}_{-0.49}$~[2]} & {\bf $-$3.60$^{+0.46}_{-0.48}$~[3]} \\ \vspace{0.2cm}
    \em{11.25 -- 11.75} &                                                 &                                                   &  {\em $-$4.19$^{+0.46}_{-0.49}$~[3]} &                                                    &                                                & {\em $-$3.87$^{+0.49}_{-0.52}$~[2]} \\    \vspace{0.2cm}                                                                                                  
 \bf{11.50 -- 12.00} &                                                   &                                                   & {\bf $-$4.66$^{+0.68}_{-0.83}$~[1]} &                                                    &                                                 & {\bf $-$4.18$^{+0.64}_{-0.78}$~[1]} \\ 
          \hline \hline
\end{tabular}
\label{tab:250LF}
\tablefoot{
\tablefoottext{a}{Values in parentheses correspond to luminosity bins that might be affected by incompleteness due to survey limits.}\\
\tablefoottext{b}{The bold (or alternatively italic) fonts denote independent luminosity bins.}}
\end{table*}
For comparison, we over-plotted previous results from the literature
at 250\,$\mu$m, specifically the LFs derived by \citet{koprowski17} from the SCUBA-2 S2CLS survey and by \citet{lapi11} from 
the {\em Herschel}-ATLAS survey. 

In the common redshift intervals, our data are almost complementary to the literature data, mostly covering the faint end of the LFs (below the knee), while
LF data from both \citet{koprowski17} and \citet{lapi11}  cover the bright end (above the knee). 
In three of the four redshift intervals in common with \citet{koprowski17} (i.e. 0.5--1.5, 1.5--2.5, and 3.5--4.5), 
in the very limited common range of luminosity our 250-$\mu$m LFs are
consistent with the SCUBA-2 one around the knee (at $z$$=$2.5--3.5, our knee is at brighter luminosities). 
Our LFs at the faint end are flatter than the extrapolation of the \citet{koprowski17} fit at low $z$, 
consistent at $z$$\sim$3, and higher at $z$$\sim$4.
In fact, in the higher redshift bin in common (3.5$<$$z$$<$4.5), our data, which reach an order of magnitude fainter luminosities, are slightly higher than the faint-L extrapolation of their Schechter fit. 
We note that \citet{koprowski17} can constrain the Schechter curve with data (from \citealt{dunlop17}) at $L_{\rm 250 \mu m}$$<$10$^{11}$ L$_{\odot}$ only in the 1.5$<$$z$$<$2.5 redshift interval. 
%Moreover, our data can be affected by some incompleteness at low luminosities and low redshifts, due to our small survey area. 

Given the error bars of our LFs, in the overlapping redshift range (i.e. at $z$$<$3.5) we are fully consistent with the {\em Herschel} LFs by \citet{lapi11}, although 
only our highest luminosity bin is in common with their faintest one. However, in the redshift
bin where our redshift distribution peaks and our data cover
a wider luminosity range (i.e. 2.5$<$$z$$<$3.5), our LF is higher than the S2CLS one at bright luminosities (e.g. at $L_{\rm 250}$$>$10$^{11}$ L$_\odot$), while it shows good agreement with the \citet{lapi11} H-ATLAS derivation (although it is almost complementary and was calculated in somewhat different redshift bins). 
Both our LFs and the {\em Herschel} ones indicate a more prominent bright end (i.e. more
luminous sources) than derived from SCUBA-2. The consistency
between our 250-$\mu$m LFs and the \citet{lapi11} ones (derived from a different sample and instrument, using a different
template SED to fit the data and a far-IR based method to derive
photometric redshifts) gives us confidence that, at least in the common redshift intervals, the photo-$z$ uncertainties do not
significantly affect our computation. 

On the other hand, the underestimated bright end by the \citet{koprowski17} S2CLS LF had previously been noted by \citet{gruppioni19}
regarding the total IR LF (obtained with the same SCUBA-2 data used for the 250-$\mu$m derivation) at $z$$=$2--3, and likely ascribed to the use of different template SEDs 
(i.e. they considered a low temperature SED, T$\simeq$35 K) to compute the 8--1000 $\mu$m luminosity as well as to incompleteness issues. 
A similar difference is now also observed with our monochromatic derivation at similar redshifts, although these are the redshifts at which our data 
are less affected by SED extrapolations. Therefore, it is likely that incompleteness issue in S2CLS data are the principal cause of the observed discrepancy.

At $z$$>$4.5 no comparison data from the literature are available, with our derivation providing the first ever determination of the luminosity and density 
distribution of dusty galaxies at such high redshifts. What is really surprising is that, even excluding the three sources
at the same redshifts of the ALPINE targets, and despite the large uncertainties, at 4.5$<$$z$$<$6 there are no hints of a significant 
decrease in the volume density of dusty galaxies (i.e. in the LF normalisation) with respect to the epoch
commonly considered to have experienced major galaxy activity (i.e. cosmic noon, $z$$\sim$1--3). A comparison between the LFs obtained with and without the three sources
is shown in the next section for the total IR LF.
%----------------------------------------------------------------- 
  \begin{figure*}
   \centering
    \includegraphics[width=18cm,height=17cm]{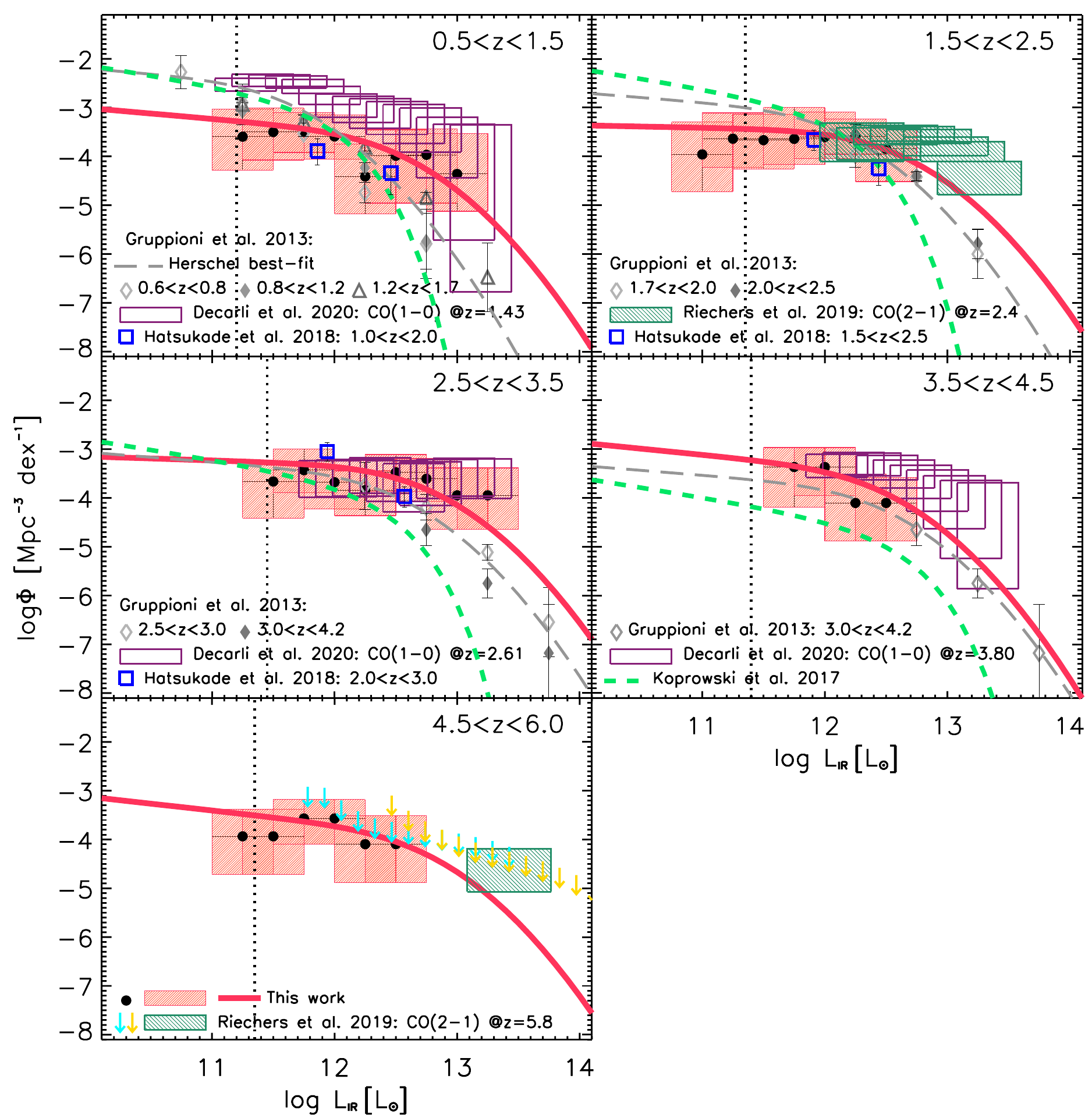}
      \caption{Total IR LF of ALPINE non-target continuum detections (red boxes and black filled circles). The luminosity bins have a width of 
   0.5 dex in ${L}_{\rm IR}$, and step through the luminosity range in steps of 0.25 dex. For this reason, the individual bins are not statistically independent. 
   The red filled boxes and error bars indicate the 1$\sigma$ errors derived through simulations (taking into account the photometric redshift uncertainties). 
   The red solid curve is the best fit modified Schechter function derived through the MCMC  analysis (See Section~\ref{sec:LFall}), while the grey long-dashed curve represents the best fit (modified 
   Schechter function) to the {\em Herschel} PEP$+$HerMES total IR LF by \citet{gruppioni13} (interpolated to the redshift bins considered in this work). 
   The {\em Herschel} PEP$+$HerMES 1/V$_{\rm max}$ data and error bars  (at slightly different redshift intervals) are plotted as grey symbols. 
   The green short-dashed curves represent the SCUBA-2 S2CLS derivation by \citet{koprowski17}. The blue open squares show the ALMA ASAGAO LFs by \citet{hatsukade18}. 
   The dark green dashed boxes and downward-pointing arrows are the COLDz CO(2-1) and CO(1-0)  LFs and limits by \citet{riechers19} at $z$$=$2.4 and 5.8, respectively, converted to $L_{\rm IR}$ as described in the text.  
   The vertical dotted line shows the ALPINE continuum survey completeness limit in $L_{\rm IR}$, computed by considering
the nominal 860-$\mu$m limiting flux of our survey (0.3 mJy (\citealt{bethermin20})) and the template, among those of the library fitting our SEDs, which provided the brightest luminosity
at the redshift of the bin.}
                 \label{figLFbol}
    \end{figure*}
%----------------------------------------------------------------- 
%----------------------------------------------------------------- 
\begin{table*}
 \caption{ALPINE total IR LF}
\begin{tabular}{ccccccc}
\hline \hline \\
 log(L$_{\rm IR}$/L$_{\odot}$) & \multicolumn{6}{c}{log($\phi$/Mpc$^{-3}$ dex$^{-1}$)   [N$_{\rm obj}$]}\\ \\
                                               & 0.5$<$$z$$<$1.5  & 1.5$<$$z$$<$2.5 & 2.5$<$$z$$<$3.5 & 3.5$<$$z$$<$4.5 & \multicolumn{2}{c}{4.5$<$$z$$<$6.0}   \\ 
                                               &   &  &  &  & No [C~II] emitters & All    \\ 
(1) & (2) & (3)  & (4) & (5)     & (6)    & (7)        \\   \hline 
 \\ \vspace{0.2cm}
   {\bf 10.75 -- 11.25\tablefootmark{a}} &                                            &     {\bf ($-$3.96$^{+0.66}_{-0.76}$~[1])\tablefootmark{b}} &  & \\ \vspace{0.2cm}
   {\em 11.00 -- 11.50} &  {\em ($-$3.60$^{+0.56}_{-0.68}$~[2])~}  &  {\em ($-$3.64$^{+0.53}_{-0.59}$~[2])~}  &                                                      &        &   {\em ($-$3.93$^{+0.55}_{-0.78}$~[1])} & {\em ($-$3.96$^{+0.55}_{-0.78}$~[1])~} \\  \vspace{0.2cm}
 {\bf 11.25 -- 11.75} & {\bf $-$3.50$^{+0.49}_{-0.57}$~[3]} &   {\bf $-$3.67$^{+0.53}_{-0.59}$~[2]}  & {\bf ($-$3.66$^{+0.37}_{-0.75}$~[2])~}  &          &  {\bf $-$3.93$^{+0.55}_{-0.78}$~[1]} & {\bf $-$3.96$^{+0.55}_{-0.78}$~[1]}\\ \vspace{0.2cm}
   {\em 11.50 -- 12.00} &  {\em $-$3.46$^{+0.40}_{-0.42}$~[4]}  &  {\em $-$3.64$^{+0.64}_{-0.52}$~[3]} & {\em $-$3.43$^{+0.43}_{-0.47}$~[4] } & {\em $-$3.37$^{+0.40}_{-0.82}$~[2] }& {\em $-$3.57$^{+0.39}_{-0.53}$~[2]  } & {\em $-$3.60$^{+0.41}_{-0.52}$~[2]} \\ \vspace{0.2cm}
 {\bf 11.75 -- 12.25} & {\bf $-$3.54$^{+0.45}_{-0.46}$ ~[3] }&  {\bf $-$3.60$^{+0.52}_{-0.44}$~[5] } & {\bf $-$3.68$^{+0.45}_{-0.55}$~[4] } & {\bf  $-$3.37$^{+0.40}_{-0.58}$~[2] } & {\bf $-$3.57$^{+0.39}_{-0.53}$~[2] } & {\bf $-$3.60$^{+0.41}_{-0.52}$~[2]}\\ \vspace{0.2cm}
    {\em  12.00 -- 12.50} &  {\em $-$4.41$^{+0.66}_{-0.76}$~[1]} &  {\em $-$3.59$^{+0.36}_{-0.37}$~[7]}  & {\em $-$3.84$^{+0.39}_{-0.52}$~[5] } & {\em  $-$4.10$^{+0.59}_{-0.78}$~[1]} &  {\em $-$4.09$^{+0.58}_{-0.79}$~[1]}  & {\em $-$4.12$^{+0.59}_{-0.78}$~[1]} \\ \vspace{0.2cm}
{\bf 12.25 -- 12.75} & {\bf $-$3.99$^{+0.53}_{-0.97}$~[2]} &  {\bf $-$3.84$^{+0.42}_{-0.67}$~[4] } & {\bf $-$3.47$^{+0.36}_{-0.39}$~[9] } & {\bf  $-$4.10$^{+0.52}_{-0.78}$~[1] }& {\bf  $-$4.09$^{+0.58}_{-0.79}$~[1] } & {\bf $-$4.12$^{+0.59}_{-0.78}$~[1]} \\ \vspace{0.2cm}
  {\em 12.50 -- 13.00 }& {\em $-$3.97$^{+0.53}_{-0.99}$~[2]}  &        &  {\em $-$3.61$^{+0.41}_{-0.46}$~[6] }&           &        & {\em $-$4.17$^{+0.59}_{-0.78}$~[1]} \\ \vspace{0.2cm}
{\bf 12.75 -- 13.25} &  {\bf $-$4.36$^{+0.83}_{-0.76}$~[1]} &       & {\bf $-$3.95$^{+0.56}_{-0.69}$~[2] }&      &   & {\bf $-$3.87$^{+0.43}_{-0.62}$~[2]} \\  \vspace{0.2cm}                                   
 {\em 13.00 -- 13.50 }&                                                            &         &   {\em $-$3.95$^{+0.56}_{-0.69}$~[2] }&  &   & {\em $-$4.17$^{+0.59}_{-0.76}$~[1]} \\     \hline \hline
\end{tabular}
\label{tab:totIRLF}
\tablefoot{
\tablefoottext{a}{The bold (or alternatively italic) fonts denote independent luminosity bins.}\\
\tablefoottext{b}{Values in parentheses correspond to luminosity bins that might be affected by incompleteness due to survey limits.}
}
\end{table*}
%----------------------------------------------------------------- 
%----------------------------------------------------------------- 
\subsection{The total infrared luminosity function}
\label{sec:totLF}
In order to derive the total IR luminosities (and LFs), we integrated the best fit SED of each source over the 8$\leq$$\lambda_{\rm rest}$$\leq$1000\,$\mu$m
($L_{\rm IR}$$=$$L$[8--1000\,$\mu$m]) range. This integration for most of our sources was performed on well-constrained SEDs covered by data in several
bands (see Figure~\ref{fig:SEDs}), while for a few sources, an extrapolation of the SED with no data constraining the far-IR peak 
was required (thus reflecting large uncertainties in $L_{\rm IR}$). 
We computed the total IR LFs in the same redshift bins considered for the monochromatic LFs at 250 $\mu$m (i.e. 0.5$<$$z$$<$1.5, 1.5$<$$z$$<$2.5, 2.5$<$$z$$<$3.5, 3.5$<$$z$$<$4.5, and 4.5$<$$z$$<$6), 
and we used the same method (1/V$_{max}$) described in the previous section.  

As already mentioned, we studied the impact of redshift uncertainties on the total IR LFs by performing a set of Monte Carlo
simulations. We iterated the computation of the monochromatic and total IR LFs 100 times, each time varying the photometric redshift of each source
(i.e. assigning a randomly selected value, according to the probability density function, PDF, and distribution associated with each redshift). 
Each time, we then recomputed the monochromatic 
and total IR luminosities, as well as the $V_{max}$, but keeping the previously found best fitting template for each
object (i.e. we did not perform the SED fitting again, since the effect of the k-correction is not significant in the sub-mm
wavelength range). For the total uncertainty in each luminosity bin, we assumed the larger dispersion between that provided by 
the Monte Carlo simulations and the Poissonian one (following \citealt{gehrels86}), although the effect of the photometric redshift uncertainty 
on the error bars is larger than the simple Poissonian value in the majority of cases.

The values of our ALPINE total IR LFs in each redshift and luminosity bin, 
with uncertainties derived by the Monte Carlo simulations, are reported in Table~\ref{tab:totIRLF}. 
The alternately independent luminosity bins are shown in italics and bold face, as in Table~\ref{tab:250LF}.
%----------------------------------------------------------------- 
%----------------------------------------------------------------- 

\subsubsection{Comparison with previous results from the literature}
\label{sec:comparison}
In Figure~\ref{figLFbol}, the total IR LFs obtained from the ALPINE sample are shown and compared with other derivations available in the literature
at similar redshifts. The {\em Herschel} (e.g. \citealt{gruppioni13}), SCUBA-2 (e.g. \citealt{koprowski17}), and ALMA (e.g. \citealt{hatsukade18}) LFs are reported in
the common or similar redshift ranges. 

We stress that this is the first total IR LF derivation reaching such faint luminosities and high redshifts: thanks to ALMA and the depth reached by the ALPINE survey,
we are finally able to sample IR luminosities typical of 'normal' (i.e., main-sequence) star-forming galaxies, rather than only those of extreme starbursts. 
We are therefore able to shape the LFs over a large luminosity range, by joining the ALMA data to the somewhat complementary {\em Herschel} and SCUBA-2 ones, 
at least up to $z$$\simeq$4. Globally, data from different surveys and wavelengths agree relatively well over the common $z$-range (up to $z$$\lesssim$4--4.5):
despite the large redshift and SED extrapolation uncertainties, the total IR LF derived from the ALPINE data is in broad agreement with those obtained in previous works. 
No continuum survey data are available for comparison at $z$$>$4.5, since our IR LF is the first at such high redshifts. We can only compare our data with line LFs 
at those redshifts.
%while the ALPINE
%survey can be undersampled in the lower $z$-bin, likely due to the small area covered.

We observe a difference with previous data in the lower redshift bin, 0.5--1.5, where both the {\em Herschel} and SCUBA-2 LFs are higher at the faint end and 
lower at the bright end, with their knee occurring at slightly fainter $L_{\rm IR}$. Indeed, the low-$L_{\rm IR}$ discrepancy (i.e. at $<$10$^{11.5}$ L$_{\odot}$) with {\em Herschel} 
is mostly determined by a single {\em Herschel} data point below the completeness limit of the ALPINE survey.
The {\em Herschel} data beyond that limit are consistent within the errors of the ALPINE derivation. 
The SCUBA-2 curve is a low-luminosity extrapolation, if we consider Figure~3 of \citet{koprowski17}. 

The faint-end extrapolations of the {\em Herschel} and SCUBA-2 LFs are still slightly steeper (and higher) than ours (at 1.5$<$$z$$<$2.5), though at those redshifts the inconsistency is also observed mostly 
below the ALPINE completeness limit, in a range where no {\em Herschel} (and probably also SCUBA-2, if we judge from the 250-$\mu$m
data points in their Figure~3) data are available to constrain the slope. 

In the luminosity range 11.5$<$log($L_{\rm IR}$/${\rm L_{\odot}})$$<$12.5, the agreement between {\em Herschel} and ALPINE is reasonably good, while at larger luminosities the ALPINE 
LF seems to remain higher (at least in the two brighter bins). The ALMA LFs from the ASAGAO survey (\citealt{hatsukade18}) agrees within the errors with our derivation (in the common luminosity range, around the knee $L^*$), 
at all redshifts (from $z$$=$0.5 to $z$$=$3.5). 

At log($L_{\rm IR}$/${\rm L_{\odot}})$$>$12.5, the S2CLS LF (\citealt{koprowski17}) shows an even steeper and lower bright end than the {\em Herschel} one, although we can only compare it 
to the best fit curve, with no data values available to check whether the agreement could have been better if we had limited our study to the luminosity range sampled by the SCUBA-2 data. 
The discrepancy with the S2CLS LF at the bright end is observed in all the common redshift bins, up to the 3.5$<$$z$$<$4.5 interval.

On the contrary, the agreement between ALPINE and the {\em Herschel} LF derivation increases with increasing redshifts, with the {\em Herschel} data being almost complementary in
luminosity, but consistent with our data within the errors in most of the common $L_{\rm IR}$ bins. 
We note that at 2.5$<$$z$$<$3.5, which is the redshift range corresponding to the peak of our $z$-distribution, the ALPINE LF seems to remain slightly higher at the bright end than the
{\em Herschel} one, while the faint end is in good agreement with the {\em Herschel} best fit extrapolation. 

At 3.5$<$$z$$<$4.5, the ALPINE data are totally complementary to the {\em Herschel} ones, the former covering the faint end and the latter the bright end of the LF, in a sort of continuity and agreement
between the two derivations. The S2CLS LF, instead, is lower than the ALPINE and {\em Herschel} ones, not only at the bright-end,
but also in normalisation and over the whole luminosity range. The underestimation of the bright end and normalisation of the total IR LF by the S2CLS data could be attributed to the 
method of deriving $L_{\rm IR}$ by \citet{koprowski17} and to an incompleteness issue due to the SCUBA-2 data sensitivity, as discussed by \citet{gruppioni19}.

The occurance of the bright end remaining significantly high, and even up to brighter luminosities than those sampled by our data, is also observed in the CO LFs by \citet{riechers19} and \citet{decarli20}, which is shown in Figure~\ref{figLFbol}
by the dark green dashed boxes and downward-pointing arrows (upper limits), and as empty purple boxes, respectively. 
These CO LFs were obtained from the blind CO surveys, CO Luminosity Density at High Redshift (COLDz; \citealt{riechers19}) and Wide ASPECS (\citealt{decarli20}), at $z$$\simeq$2.4, 5.8 and $z$$=$1.43, 2.61, 3.80, respectively.
In order to allow a direct comparison with our data, the CO luminosities ($L'_{CO}$, in K km s$^{-1}$ pc$^2$) were converted to IR luminosities (in L$_{\odot}$) according to \citet{carilli13} to pass from $L'_{\rm CO(1-0)}$ to 
$L_{\rm IR}$ (i.e. log$L_{\rm IR}$$=$1.37 log$L'_{\rm CO(1-0)}$$-$1.74), and \citet{decarli16} to convert $L'_{\rm CO(2-1)}$ to $L'_{\rm CO(1-0)}$ (i.e., log$L'_{\rm CO(1-0)}$$=$log$L'_{\rm CO(2-1)}$$-$log(0.76)}).
We note that in the common luminosity bins, the COLDz derivation is in very good agreement with our estimate, with the CO LFs extending the high bright end to even higher luminosities. 
Indeed, the recent finding that in the COLDz survey there may be an overdensity at $z$$\sim$5 capturing three bright dusty starbursts (\citealt{riechers20}) could partially explain such a high CO LF at 4.5$<$$z$$<$6 (see e.g. our LF obtained by including the two [C~II] emitters likely associated with the ALPINE targets shown in Figure~\ref{figLF45_6} and discussed in the next section).
The ASPECS LF is also in agreement with our estimate, especially at 2.5$<$$z$$<$3.5, while at 3.5$<$$z$$<$4.5 it extends the bright end to higher luminosities than are sampled by our data. 
At low redshift (i.e. 0.5$<$$z$$<$1.5),
it is well consistent with our LF at the bright end, while it is higher at fainter luminosities (i.e. $<$10$^{12}$ L$_{\odot}$).
Overall, the good consistency with these completely independent derivations validate the existence of a prominent bright end in the dusty galaxies' LFs, 
which has so far been highly debated in the literature and often attributed to source blending due to
low resolution in far-IR/sub-mm data.
%----------------------------------------------------------------- 
%----------------------------------------------------------------- 
  \begin{figure*}
   \centering
 \includegraphics[width=18cm]{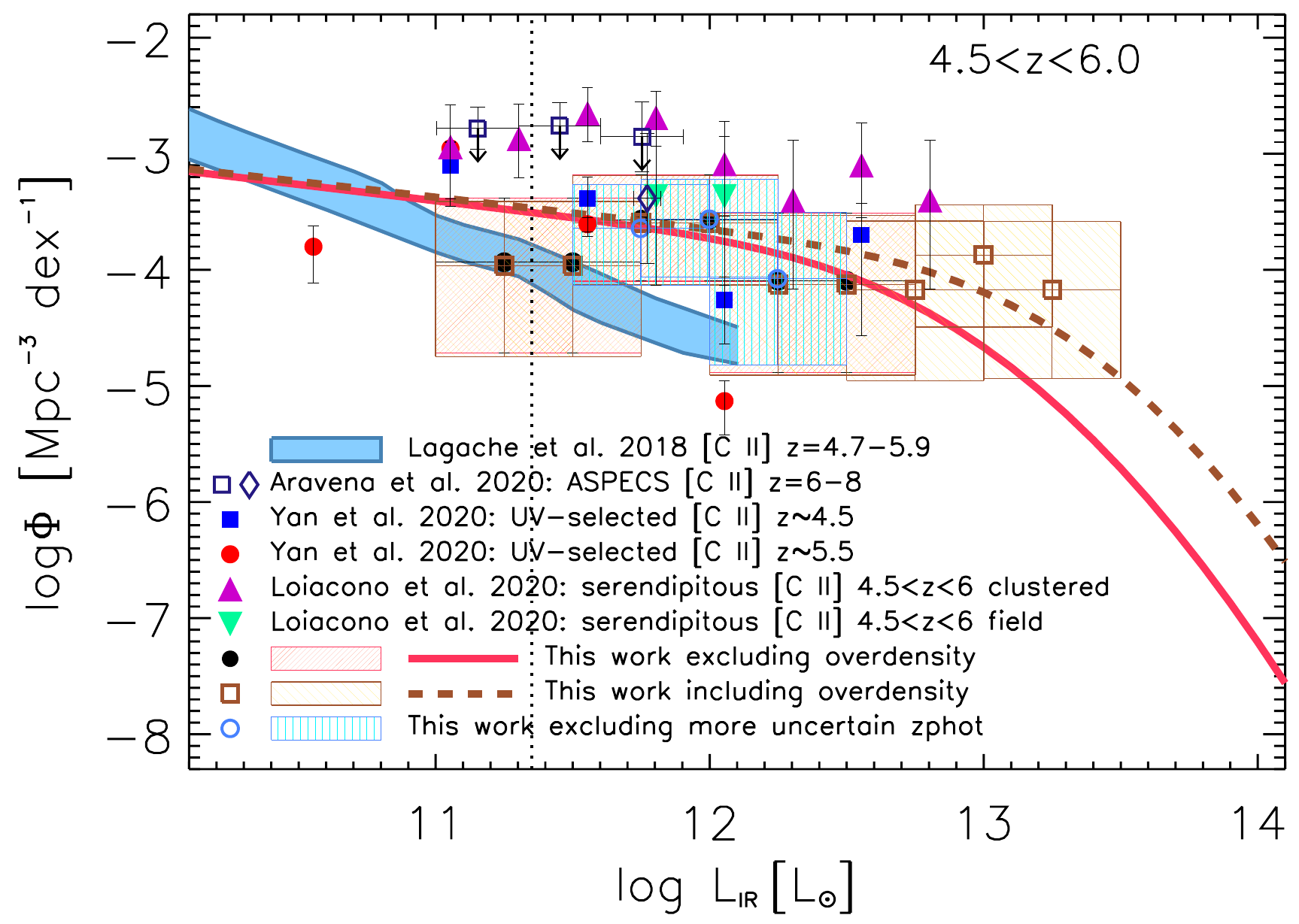}
    \caption{Total IR LF of ALPINE non-target continuum detections in redshift interval 4.5$<$$z$$<$6: the results shown in Figure~\ref{figLFbol} (red boxes and black filled circles, red solid curve) -- obtained by excluding the sources with spectroscopic redshift equal to that of the ALPINE target at the centre of the pointing 
    (i.e. two sources with optical/near-IR identification) -- are compared to those obtained by also including these objects (yellow boxes and brown open squares). 
   The brown dashed line is the MCMC modified Schechter fit to the latter LF derivation. The cyan dashed boxes show the LF recomputed after excluding the source with
more uncertain photo-$z$ (the one at $z$$=$5.85). This test was performed to check the robustness of our result at these critical redshifts.
   The error bars in all the LFs show the 1$\sigma$ errors obtained by combining the Poissonian errors with those derived with simulations, the latter considering the 
   photometric redshift uncertainties. The vertical dotted line shows the ALPINE continuum survey completeness limit in this redshift interval.  For comparison, we report the ALPINE [C~II]158 $\mu$m LFs (converted to total IR LFs as described in the text) at similar redshifts, obtained by \citet{yan20} for the UV-selected ALPINE targets detected in 
   [C~II] ($z$$\simeq$4.5: blue filled squares, $z$$\simeq$5.5: red filled circles), and by \citet{loiacono20} for
   the serendipitous [C~II] detections at 4.5$<$$z$$<$6.0 (lines falling in the same spectral window of the targets, that is, `clustered': violet filled triangles; lines separated by that of the targets by 
   $>$2000 km s$^{-1}$, that is, `field': green upside-down triangles). The [C~II] LF model predictions by \citet{lagache18a} at $z$$=$4.7--5.9 are reported via the light blue coloured area ($z$$=$4.7 upper boundary, $z$$=$5.9 lower), while
   the ASPECS derivation of the [C~II] LFs at $z$$=$6--8 by \citet{aravena20} are shown as dark blue open symbols (squares with downward-pointing arrows show upper limits, and the diamond shows the lower limit assuming that only one source is real).}
              \label{figLF45_6}
    \end{figure*}
%----------------------------------------------------------------- 
%----------------------------------------------------------------- 
\subsubsection{Luminosity function at $z$$\simeq$5}
\label{sec:LFz5}
In the highest redshift bin covered by our survey (4.5$<$$z$$<$6), we find no comparison data in the IR from the literature, but only constraints from the CO emission (\citealt{riechers19}).
The hints provided by our LF in the $z$$=$4.5--6 redshift range, in good agreement with those by \cite{riechers19}, are that the volume density of dusty sources 
remains high (almost as much as at $z$$\simeq$2--3), with no significant drop in normalisation at $z$$>$2.5--3. The global shape of the LF does not 
change significantly from low to high redshift. The faint end of the LF does not show any evident
steepening, and the LF knee, though barely constrained by
data, seems to fall at bright luminosities, similarly to those found at lower redshifts.

In Figure~\ref{figLF45_6}, we compare the total IR LF at 4.5$<$$z$$<$6 obtained
by excluding the sources found at the same redshift of
the ALPINE targets (the same is shown in Figure~\ref{figLFbol}: red boxes
and black filled circles) to that obtained by also including two of these
galaxies (i.e. those with an optical/near-IR identification, which are shown by yellow dashed boxes and brown open squares). We note that
the inclusion of the two [C~II] emitters does not alter the shape of
the LF in the common luminosity range. What actually happens
is that these sources populate the higher luminosity bins, extending
the bright end of the LF to higher $L_{\rm IR}$. The reason why these
sources (associated with the ALPINE central targets) have
luminosities higher than the other sources at similar redshifts is
not clear: however, this investigation is beyond the scope of this
work and will be considered in a future paper.

We also performed a further check to test the robustness of our result
in this redshift bin by recomputing the LF after excluding the source with
more uncertain photo-$z$, meaning the one at $z$$=$5.85.
The result is shown in Figure~\ref{figLF45_6} with cyan dashed boxes. 
The luminosity range covered by the LF is smaller (i.e. the excluded source
was populating the lower luminosity bin, and was partially affected by incompleteness), 
but the normalisation remains exactly the same and also the best fit curve 
passes through the data well.
We therefore find that even if this source were at a smaller than estimated redshift , our high-$z$ LF derivation and 
our conclusions would not be affected.

For comparison, in the figure we plot also the [C~II] LFs obtained
at similar redshifts by \citet{yan20} for the UV-selected
ALPINE targets detected in [C~II] ($z$$\simeq$4.5: blue filled squares,
$z$$\simeq$5.5: red filled circles), and by \citet{loiacono20}
for the serendipitous [C~II] detections in the ALPINE
pointings (at 4.5$<$$z$$<$6.0). The latter LF is divided in two derivations:
one considers the lines in the same ALMA spectral window
of the targets (i.e. 'clustered': violet filled triangles), and the
other the lines spectrally distant from the targets by $>$2000 km
s$^{-1}$ (i.e. 'field': green upside-down triangles). To allow the
comparison with our continuum data, we converted the
[C~II] luminosities ($L_{\rm [C~II]}$) to $L_{\rm IR}$ by following the recipe
of \citet{hemmati17} by adopting log$_{\rm 10}$($L_{\rm FIR}$/$L_{\rm [C~II]}$)$=$2.69
(value from \citealt{zanella18}),
then a ratio $L_{\rm IR}$/$L_{\rm FIR}$($=$$L_{\rm [8-1000\mu m]}$/$L_{\rm [42-122\mu m]}$)$=$1.3. 
The results do not change if we convert $L_{\rm [C~II]}$ to SFR using the 
\citet{delooze14} relation, then the SFR to $L_{\rm IR}$ through the \citet{kennicutt98} calibration.

The [C~II] LFs of the ALPINE targets (UV-selected; \citealt{yan20}) at both $z$$\sim$4.5 and 5.5 are lower and steeper than 
our best fit curve, although the high-L data point at $z$$\sim$4.5, at log$_{\rm 10}$($L_{\rm IR}$/L$_{\odot}$)$=$12.5, rises again and reaches our values.
The fact that the ALPINE targets were selected in the UV rest frame can explain the steeper bright end, because the UV selection can miss the dustier sources.
%The converted [C~II] luminosities reach $L_{\rm IR}$ below our completeness limit and cover a fainter luminosity range than our continuum data. 

On the other hand, the [C~II] LF of the field
serendipitous detections (\citealt{loiacono20}) is in perfect
agreement with our data. The [C~II] LF of
the clustered serendipitous detections instead is slightly higher
than our derivation (though consistent within the uncertainties), especially below our completeness limit.
Similarly to our LF obtained by including the two sources at the redshift of the ALPINE targets, the
[C~II] clustered LF also extends to higher luminosities than the field one. This seems to imply that
sources belonging to an overdensity are more luminous than the field ones.

The model prediction of [C~II] LF by \citet{lagache18a} at $z$$=$4.7--5.9 is also reported (as a light blue coloured area: $z$$=$4.7 upper boundary, $z$$=$5.9 lower)
in Figure~\ref{figLF45_6}: it is generally steeper than our LF, slightly lower at bright luminosities, and consistent at low $L_{\rm IR}$ (though we are likely affected by incompleteness
in the first two luminosity bins).
The ASPECS [C~II] LF at $z$$=$6--8 by \citet{aravena20}, converted to $L_{\rm IR}$ by means of the same method used for \citet{loiacono20} and \citet{yan20} LFs, is shown with dark blue open symbols 
(the squares with downward-pointing arrows show the upper limits, and the diamond shows the conservative value obtained by assuming that only the source with an optical counterpart is real). The 
`conservative' ASPECS data point is in very good agreement with our LF, with the latter never exceeding their upper limits.

%----------------------------------------------------------------- 
%----------------------------------------------------------------- 
\subsubsection{Evolution}
\label{sec:LFall}
\begin{figure}
\includegraphics[width=9.cm]{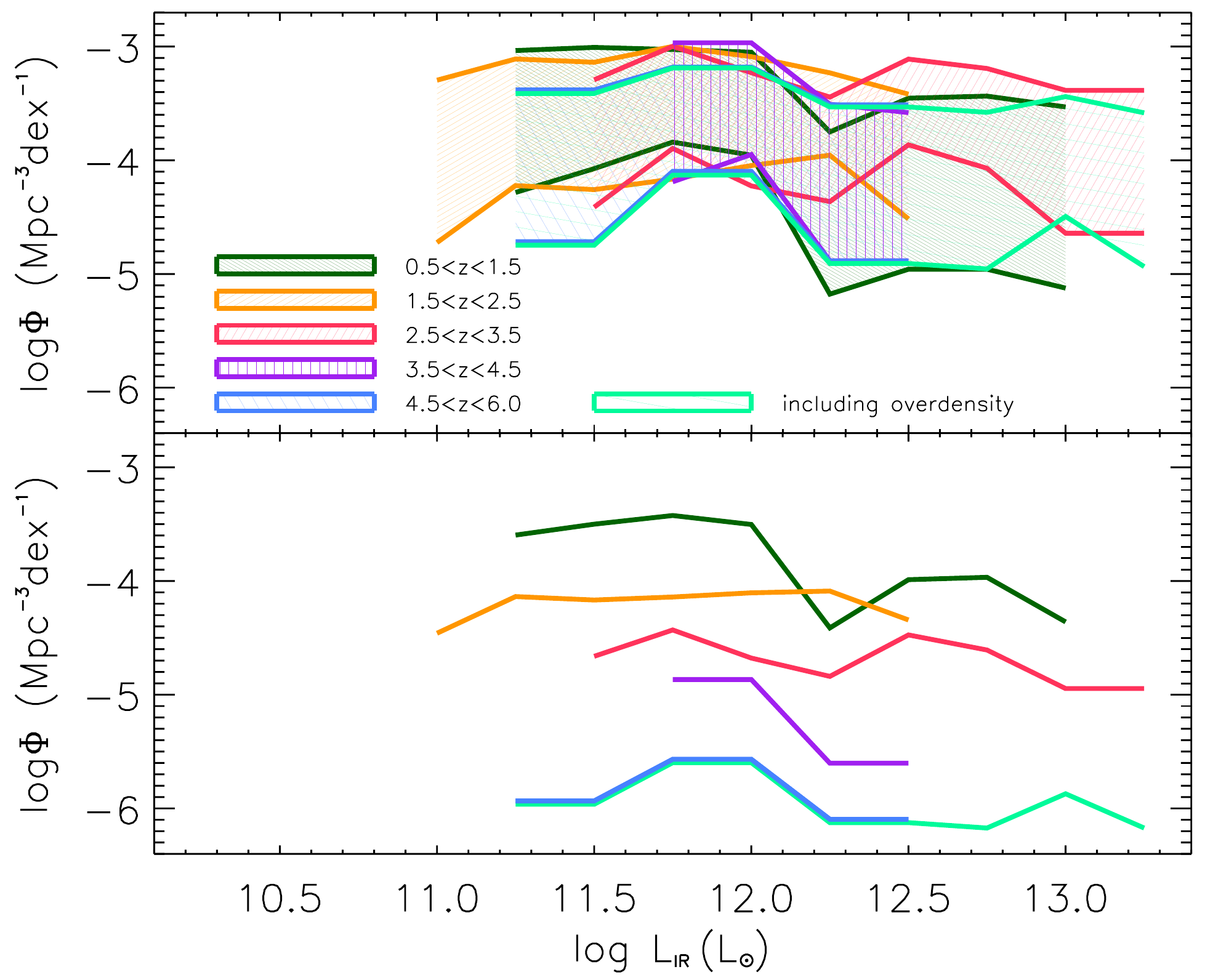}
\caption{Total IR LF shown in the different panels of Figure~\ref{figLFbol} plotted at all the different redshift intervals considered in this study, from $z$$\sim$0.5 to $z$$\sim$6. 
$Top$: the different coloured areas represent the $\pm$1$\sigma$ uncertainty regions at different redshifts obtained from the Monte Carlo simulations. 
We note that for the highest redshift interval (4.5$<$$z$$<$6.0), in both panels the LF is shown for both derivations. This was obtained by excluding (blue) and including (green) the sources at the same 
redshifts of the ALPINE targets (see legend).
$Bottom$: median value of the LFs in each luminosity bin at all the redshift intervals, each scaled by a factor of 0.5 relatively to the previous one, from the lowest to the highest redshift. The different colours show the different redshift intervals, with the same colours of the $top$ panel. }
\label{figLFallz}
\end{figure}

In order to facilitate the comparison between the LFs at different
redshifts, in the $top$ panel of Figure~\ref{figLFallz} we plot the total IR LFs at all redshifts with their $\pm$1$\sigma$
uncertainty regions (different colours for different $z$-intervals). The errors are large, therefore 
it is difficult to detect any significant evolution of the LF with $z$; 
it is however surprising to note that there does not seem to be any appreciable evolution from $z$$\sim$0.5 to $z$$\sim$6, both in shape and normalisation.

However, we must stress that with ALPINE we are mostly covering the faint end of the total IR LF over the whole redshift range, with the exception of the 2.5$<$$z$$<$3.5 interval, 
where we span a slightly larger range of luminosities and we are also able to reach luminosities above the knee. 
Therefore, the apparent non-evolution of the LF found in this work is not inconsistent with previous results (i.e. based on {\em Herschel} data) claiming a strong luminosity evolution up 
to $z$$\simeq$2--3 (e.g. \citealt{gruppioni13}). This is because the evolution in the {\em Herschel} LFs is observed principally at its bright end, where ALPINE has limited constraining power.

In the $bottom$ panel of Figure~\ref{figLFallz}, we show only the median values of the LFs in each luminosity bin at all the redshift intervals, each scaled by a factor of 0.5 relatively to the previous one,
from the lowest to the highest redshift, in order to facilitate the shape comparison.
From the figure, we note that in general the LFs seem to present two 'bumps', one at lower and the other at higher luminosities, though at very low significance (i.e. 1.5$\sigma$).
The two bumps are noticeable 
%, as if two different populations were dominating different luminosity regimes. 
in particular where our sample covers the wider range of luminosities, that is, at  $z$$=$0.5--1.5 and 2.5--3.5
(dark green and red dashed areas -- $top$ -- and curves -- $bottom$).
%where the two bumps, peaking at $L_{\rm IR}$$\simeq$10^{11.7} and 10^{12.5} L$_{\odot}$ respectively, are better shaped by data. 
In the lowest redshift bin, the 
bump at brighter luminosities has a lower normalisation than the one peaking at fainter $L_{\rm IR}$. 
At $z$$=$1.5--2.5 and 3.5--4.5 our LFs sample only the fainter luminosities (and perhaps the fainter bump), while
at 4.5--6 a sort of double-peaked distribution is observed
only when we consider all the serendipitous detections (i.e. without excluding
the sources at the same redshift of the ALPINE targets; bright green). 
%By comparing the results from the lowest to the highest redshift, the peaks of the two bumps 
%seem to shift towards higher luminosities with increasing $z$, the higher-$L$ one 
%increasing in normalisation, at least from $z$$\simeq$0.5 to 3.5.
%If this effect is real, it would imply significant luminosity evolution up to $z$$\simeq$6 for 
%the populations shaping the double LF distribution. 
If the two bumps were real and maybe due to two different populations, 
the one responsible for the higher $L_{\rm IR}$ bump
will become more dominant with increasing $z$.
We would need more data to confirm these hints: with the current data we
can only make speculations.
%that if the higher luminosity population 
%be dominated by galaxies containing an AGN, the latter becoming
%more important at higher redshifts,  the result is in agreement with the
%scenario described by \citet{gruppioni13} to $z$$\sim$3--4).
%----------------------------------------------------------------- 
%----------------------------------------------------------------- 

In general, the ALPINE total IR LFs seem to confirm the `flat' shape already found by {\em Herschel}, at both its faint and bright ends. 
In particular, the bright end remains significantly high even in the higher redshift interval, where
the volume density of ultra-/hyper-luminous IR galaxies equals that of more 'normal' galaxies.
The presence of such (and so many) bright IR galaxies at high-$z$ is a real challenge for galaxy formation models (already 
at $z$$\sim$2--3, even worse at higher $z$),
with no current model being able to explain the existence of massive, dusty, and actively star-forming galaxies at such early epochs.  

On the other hand, the flat faint end implies a minor contribution from low-luminosity and/or low-mass galaxies to
the dust emission, while a main contribution from high-luminosity and/or high-mass systems, up to the
highest redshifts, is required by the bright end.
The increasing number of low-mass systems with increasing redshift predicted by the hierarchical structure formation scenario 
is not observed in our data. It might however be that low-mass and/or low-dust-mass systems become more important or numerous at higher redshifts,
but do not produce (heat) enough far-IR emission (dust) to be detected in far-IR/mm surveys.
In order to study the evolution of the total IR LF and of the SFRD with $z$, we obtained a parametric estimate of the
luminosity function at different redshifts. Although the ALPINE LFs may have a more complex shape 
(i.e. a double-peaked distribution), for simplicity and to better compare the values of the parameters with previous results, we assumed a modified Schechter function
(i.e. \citealt{saunders90}), with $\phi(L)$ given by
\begin{equation}
\phi(L)~{\rm dlog}L=\phi^{\star}\left(\frac{L}{L^{\star}}\right)^{1-\alpha} \exp\left[-\frac{1}{2\sigma^2}\log_{10}^2\left(1+\frac{L}{L^{\star}}\right)\right]{\rm dlog}L, 
\label{eq:schechter}
\end{equation}
behaving as a power law for $L \ll L^{\star}$ and as a Gaussian in $\log L$ for $L\gg L^{\star}$.
The adopted LF parametric shape depends on four parameters ($\alpha$, $\sigma$, $L^{\star}$ and $\phi^{\star}$), whose best fitting values and uncertainties
were derived using a Markov chain Monte Carlo (MCMC) procedure. Since the ALPINE data do not sample the bright luminosities, the slope
of the bright end is almost unconstrained: we therefore fixed the value of $\sigma$ (the parameter shaping the bright-end slope) to that found for 
the {\em Herschel} LFs ($\sigma$$=$0.5).  We considered flat priors for the other three parameters ($\alpha$, $L^{\star}$ and $\phi^{\star}$), 
%the values found by minimising the Likelihood function in the independent luminosity bins. The MCMC exploration has then been limited to a reasonably wide range of values 
limiting the MCMC exploration to a reasonably wide range of values 
(i.e. log($L^{\star}$/L$_{\odot})$: [10,13], log$(\phi^{\star}$/Mpc$^{-3}$dex$^{-1})$: [$-$2,$-$5], $\alpha$: [$-$1,2]). 
The result of the MCMC analysis is shown in Figures~\ref{figLFbol} (red solid curve) and~\ref{figLF45_6} (red solid and brown 
dashed curves) and presented in Table~\ref{tab:MCMC}.
\begin{table}
\caption{MCMC best fitting parameters}             % title of Table
\label{tab:MCMC}      % is used to refer this table in the text
\centering                          % used for centering table
\begin{tabular}{c c c c}        % centered columns (4 columns)
\hline\hline         \\        % inserts double horizontal lines
$z$ & $\alpha$ & log$(L^{\star}$/L$_{\odot})$ &  log$(\phi^{\star}$/Mpc$^{-3}$dex$^{-1})$ \\    
(1) & (2) & (3)  & (4) \\ % table heading 
\hline    \\    \vspace{0.2cm}
                 % inserts single horizontal line
    0.5--1.5 & 1.22$^{+0.15}_{-0.17}$ & 11.95$^{+0.41}_{-0.36}$ & $-$3.44$^{+0.24}_{-0.23}$ \\      % inserting body of the table
 \vspace{0.2cm}
    1.5--2.5 & 1.15$^{+0.17}_{-0.12}$ & 12.01$^{+0.36}_{-0.43}$    & $-$3.45$^{+0.18}_{-0.19}$\\
  \vspace{0.2cm}
    2.5--3.5 & 1.08$^{+0.17}_{-0.11}$ & 12.12$^{+0.20}_{-0.24}$     & $-$3.32$^{+0.14}_{-0.15}$ \\
  \vspace{0.2cm}
    3.5--4.5 & 1.25$^{+0.43}_{-0.55}$ & 11.90$^{+0.65}_{-0.43}$    & $-$3.43$^{+0.49}_{-0.40}$ \\
  \vspace{0.2cm}
    4.5--6.0 & 1.28$^{+0.46}_{-0.48}$ & 12.16$^{+0.85}_{-0.76}$    & $-$3.73$^{+0.46}_{-0.54}$ \\ 
\hline     \hline                             %inserts single line
\end{tabular}
\end{table}
%----------------------------------------------------------------- 
%----------------------------------------------------------------- 
\section{Contribution to the cosmic SFRD}
\label{sec:SFRD}
%----------------------------------------------------------------- 
\begin{figure*}
\includegraphics[width=18cm]{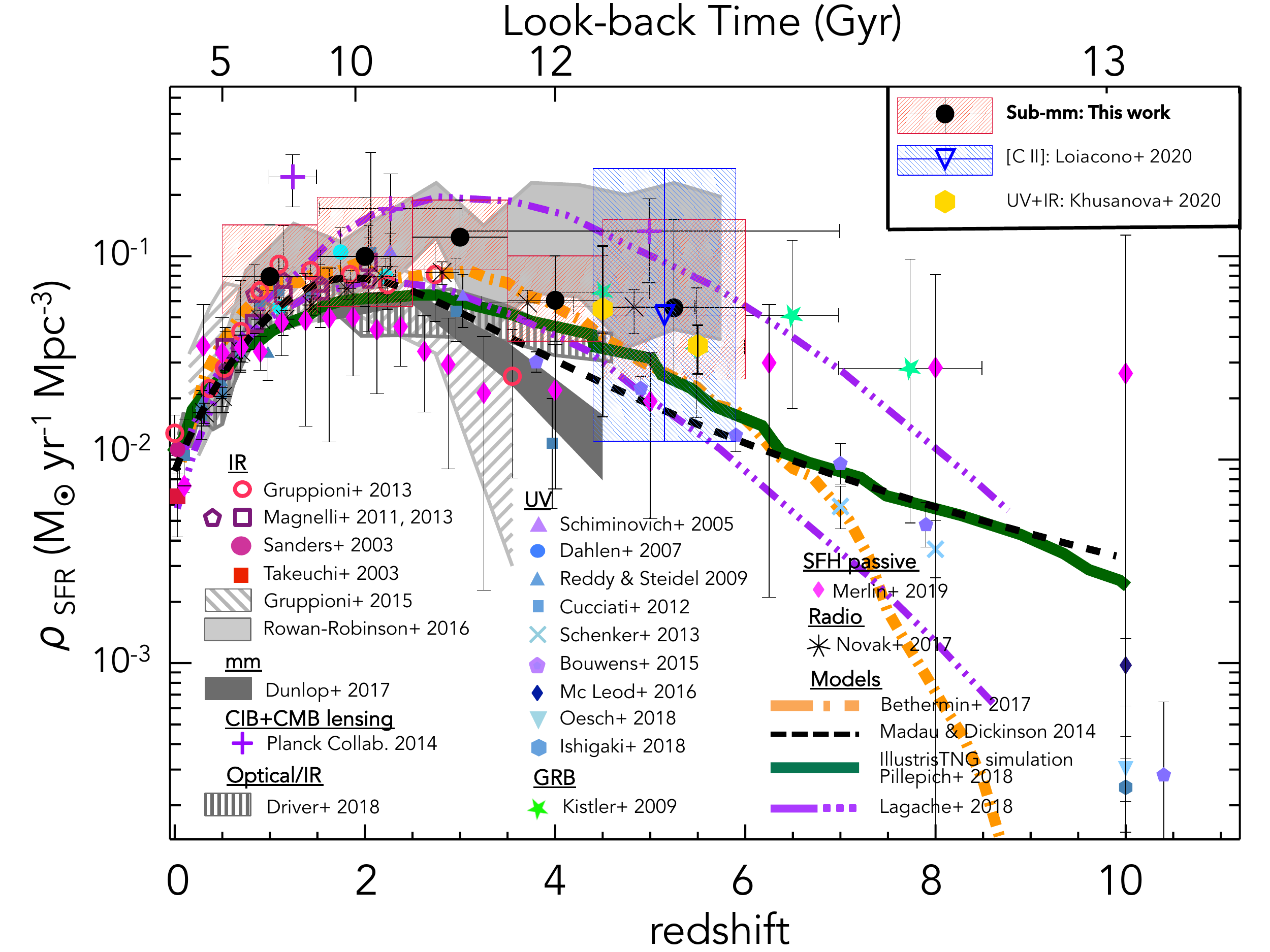}
\caption{Redshift evolution of co-moving star formation rate density ($\rho_{\rm SFR}$), obtained by integrating the modified Schechter function that best reproduces the ALPINE total IR LF 
of the continuum non-target detections (excluding the [C~II] emitters): black circles. The error bars and the red boxes around our data points show the 1$\sigma$ uncertainty range derived through the MCMC analysis of the LF. 
%The yellow box and brown open square show the SFRD estimated by integrating the best-fit curve to the 4.5$<$$z$$<$6 LF obtained from all the continuum detections, including the 5 [C~II] emitters.
The SFRD estimates from ALPINE (legend in the top-right corner of the plot) are also shown for comparison: the blue box with blue open square represents the result obtained from the [C~II] LF of the serendipitous line emitters by \citet{loiacono20}, while the
yellow filled  hexagons with error bars are the values obtained by \citet{khusanova20} from the UV$+$IR emission of the ALPINE targets.
For comparison, estimates from other surveys (UV: \citealt{schiminovich05, dahlen07, reddy09, cucciati12, schenker13, bouwens15, oesch18}; optical/near-IR: \citealt{driver18, merlin19}; far-IR: \citealt{sanders03, takeuchi03, magnelli11, magnelli13, gruppioni13, gruppioni15, rowanrobinson16}; mm: \citealt{dunlop17}; radio: \citealt{novak17}; gamma-ray bursts: \citealt{kistler09}) are also shown (grey shaded areas and open or filled symbols), as described in the legend at the bottom of the plot. 
The models by \citet{madau14} and \citet{bethermin17} are shown as black dashed and orange dot-dashed curves, respectively, while the prediction of the IllustrisTNG simulation (\citealt{pillepich18}) is shown as a dark green solid curve. Also shown are the measurements derived from the cross-correlation between the lensing map of the CMB and the CIB (light blue crosses with error bars, \citealt{planck14}) and the prediction by \citet{lagache18b} obtained by modelling the CIB (violet triple-dot-dashed curves,
showing the pessimistic and optimistic cases).  }
\label{figldz}
\end{figure*}
%----------------------------------------------------------------- 

We derived the evolution of the co-moving luminosity density ($\rho_{\rm IR}$) of the ALPINE continuum non-target sources by integrating the total IR LF in the different redshift bins, from $z$$\simeq$0.5--1.5 to $z$$\simeq$4.5--6
(i.e. $\rho_{\rm IR}(z)$$=$$\int{\phi({\rm log}L_{\rm IR},z)~L_{\rm IR}~{\rm dlog}L_{\rm IR}}$). 
In order to do this, we extrapolated the modified Schechter functions that best reproduce our data down 
to log($L_{\rm IR}$/L$_{\odot}$)$=$8.  
If the overall contribution to the IR luminosity density from the AGN components of galaxies is small, $\rho_{\rm IR}$ can be considered as a proxy of the co-moving SFRD ($\rho_{\rm SFR}(z)$), assuming the \citet{kennicutt98} relation that
connects the SFR and $L_{\rm IR}$.
% which is a crucial tool for understanding galaxy formation and evolution.
%Therefore, the total IR LF can be integrated to study the evolution of the total comoving SFRD with $z$ ($\rho_{\rm SFR}(z)$), which is a crucial tool for understanding galaxy formation and evolution. 
For the majority of our SEDs, we cannot reliably disentangle the AGN from the star formation contribution, since we do not have enough data in the mid-/far-IR. However, although we cannot exclude the
presence of an AGN inside our galaxies, we notice that the large majority or the ALPINE SEDs are best fitted by star-forming or composite templates rather than by AGN-dominated ones. 
Indeed, the best fit templates that reproduce the ALPINE SEDs are similar to those found to reproduce the majority of the {\em Herschel} PEP$+$HerMES galaxies at $z$$\simeq$2--3 (\citealt{gruppioni13}). 
Furthermore, their decomposition and separation into AGN and SF contributions (performed by \citealt{delvecchio14}) showed a negligible contribution to $L_{\rm IR}$ ($<$10 per cent) from the AGN and an SF 
component dominating the far-IR, even in the SEDs fitted by more powerful AGN templates (see also \citealt{lemaux14}).
Since in ALPINE we found very few AGN-dominated templates, we do not expect that contamination related to accretion activity can significantly affect the results in terms of 
$\rho_{\rm SFR}$. We therefore used the relation found by \citet{kennicutt98} to convert $L_{\rm IR}$ to SFR, then $\rho_{\rm IR}(z)$ to $\rho_{\rm SFR}(z)$, for a \citet{chabrier03} IMF: 
\begin{equation}
{\rm SFR (M_{\odot} yr^{-1}) \simeq 1.09 \times 10^{-10}~L_{IR} (L_{\odot})}
.\end{equation}

In Figure~\ref{figldz}, we show $\rho_{\rm SFR}(z)$ estimated from our total IR LF (values presented in Table~\ref{tab:SFRD}) and compare it with results obtained from previous surveys in different bands, from the optical/UV to the radio 
(see references in the figure legend and caption). 
Since our lower redshift bin is centred at $z$$=$1, our co-moving SFRD does not show the rapid rise from $z$$\sim$0 to $z$$\sim$1 observed in other surveys. 
It does, however, show a very flat distribution from $z$$=$0.5 to $z$$=$6, with no significant decrease beyond the cosmic noon ($z$$\simeq$1--3), as is observed from optical/UV surveys.
Other SFRD derivations from the ALPINE collaboration are shown for comparison: from the serendipitous [C~II] LF (blue box; \citealt{loiacono20}) and from the UV+IR SFR of the ALPINE targets (yellow filled hexagons;
\citealt{khusanova20}), highlighted in the top-right corner of the plot. The [C~II] result agrees well with our $z$$\simeq$5 value, and also the UV+IR target data are consistent with ours within the uncertainties, though
the higher redshift one is slightly lower (possibly due to the UV selection missing highly obscured galaxies).
\begin{table}
\caption{Star formation rate density}             % title of Table
\label{tab:SFRD}      % is used to refer this table in the text
\centering                          % used for centering table
\begin{tabular}{c c c c}        % centered columns (4 columns)
\hline\hline         \\        % inserts double horizontal lines
$z$ & SFRD  & SFRD$\_$min & SFRD$\_$max  \\    
   & \multicolumn{3}{c}{[M$_\odot$ yr$^{-1}$ Mpc$^{-3}$]} \\    
(1) & (2) & (3)  & (4)  \\  % table heading 
\hline    \\    \vspace{0.2cm}
                 % inserts single horizontal line
    0.5--1.5 & 7.93$\times$10$^{-2}$ & 5.19$\times$10$^{-2}$  & 1.42$\times$10$^{-1}$ \\      % inserting body of the table
 \vspace{0.2cm}
    1.5--2.5 & 9.96$\times$10$^{-2}$ & 5.64$\times$10$^{-2}$  & 1.94$\times$10$^{-1}$ \\
  \vspace{0.2cm}
    2.5--3.5 & 1.23$\times$10$^{-1}$ & 8.57$\times$10$^{-2}$  & 1.88$\times$10$^{-1}$ \\
  \vspace{0.2cm}
    3.5--4.5 & 6.06$\times$10$^{-2}$ & 3.82$\times$10$^{-2}$  & 1.00$\times$10$^{-1}$ \\
  \vspace{0.2cm}
    4.5--6.0 & 5.57$\times$10$^{-2}$ &  2.49$\times$10$^{-2}$  & 1.51$\times$10$^{-1}$ \\ 
\hline     \hline                             %inserts single line
\end{tabular}
\end{table}

Our data are also in very good agreement with the far-IR results (from {\em Spitzer} and {\em Herschel}) over the common redshift range (e.g. 1--3: \citealt{rodighiero10, magnelli11,magnelli13, gruppioni13}),
and in particular with the sub-mm results of \citet{rowanrobinson16} (highly debated because they are based on exceptional {\em Herschel} SPIRE 500-$\mu$m galaxies) over the whole redshift range. 
In addition, we find a good agreement 
with the results of \citet{kistler09} from gamma-ray bursts at $z$$>$4, with the measurements derived from the cross-correlation between the lensing map of the CMB and the CIB  by the \citealt{planck14}, and with the $\rho_{\rm SFR}(z)$ derived by \citet{novak17} from radio surveys at $z$$\simeq$1--5. 

On the other hand, the SFRD derived from optical/UV surveys, although extending to higher redshifts (i.e. $z$$\simeq$10), are always significantly lower than our estimates at $z$$>$3. The difference increases with redshift, reaching a factor of about 10 at $z$$\sim$6. 
When performing this comparison, we must note that, while we integrated the IR LF down to 10$^8$ L $_{\odot}$ (i.e. an SFR of $\simeq$10$^{-2}$ M$_{\odot}$ yr$^{-1}$) to derive the SFRD, the SFRD estimates for UV-selected galaxies are 
always integrated down to the detection limits of the highest redshift LF (e.g. to an SFR limit of 0.3 M$_{\odot}$ yr$^{-1}$ at z=10; \citealt{oesch18}). This is done because the faint-end slope of the UV LF at high redshift 
is found to be very steep, leading the UV LF integration to diverge. However, given the very flat faint end of our IR LF, integrating it to SFR limits similar to those of the UV works would not significantly modify our results.

\begin{figure}
\includegraphics[width=8.8cm]{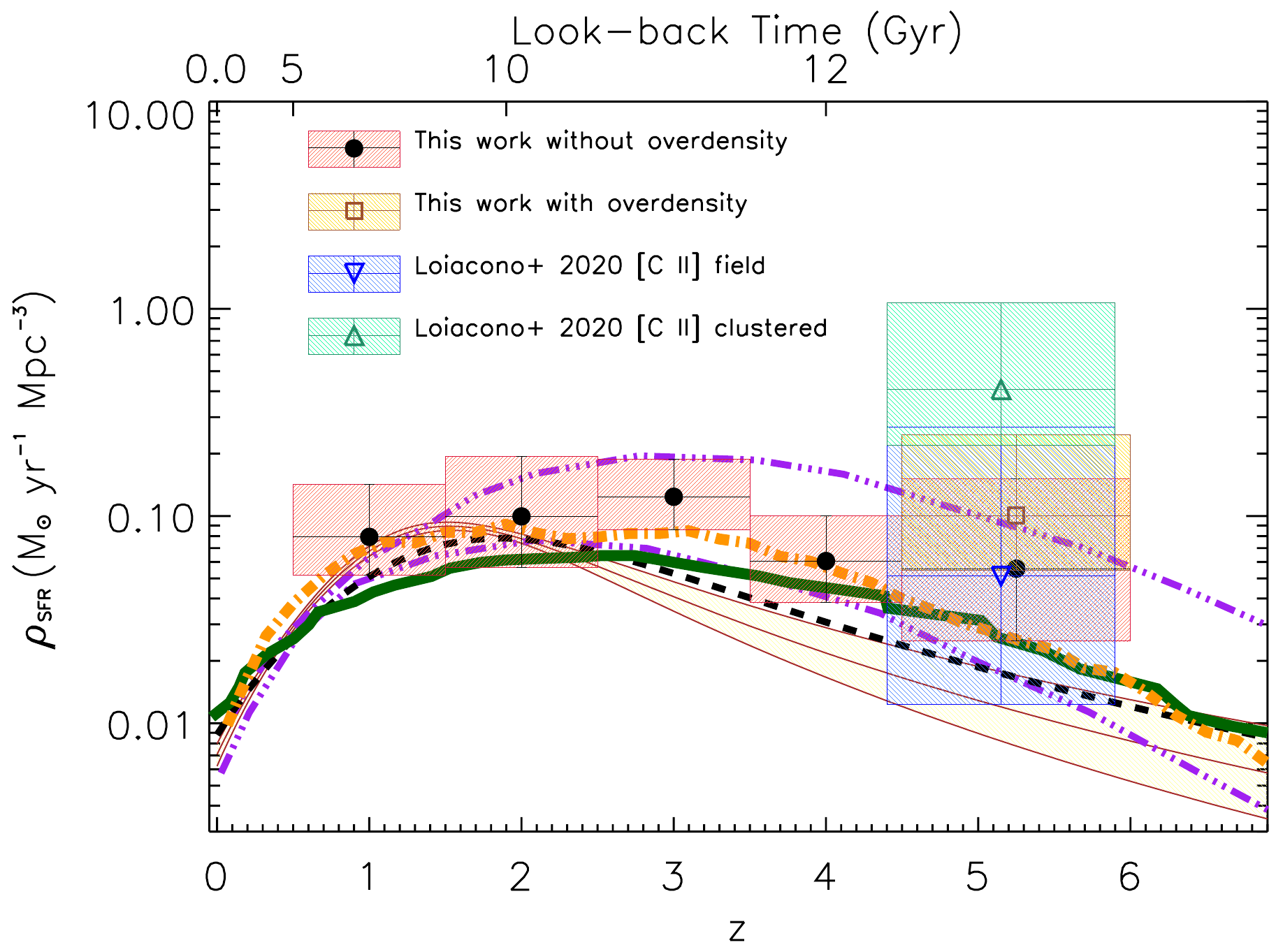}
\caption{Comparison between SFRD obtained by excluding the two [C~II] emitters with optical/near-IR counterparts (red boxes and black circles, same as in Figure \ref{figldz}) and that 
estimated by integrating the best fit curve to the 4.5$<$$z$$<$6 LF obtained from all the continuum detections, including the two [C~II] emitters 
(yellow boxes and brown open square).
For comparison, we also report the results obtained by \citet{loiacono20} by integrating the [C~II] LF of the serendipitous line emitters for field  (i.e. lines separated by that of the ALPINE targets by 
   $>$2000 km s$^{-1}$: blue box and open up-side down triangle) and 
clustered sources (i.e. lines falling in the same spectral window of the ALPINE targets: green box and open triangle).
The models by \citet{madau14}, \citet{bethermin17}, and \citet{pillepich18} are also reported as black dashed, orange dot-dashed, and dark green solid curves, respectively. In addition, we plot the predictions 
obtained by modelling the CIB of \citet{manyiar18}, as a light yellow dashed area, and by \citet{lagache18b} as violet triple-dot-dashed lines (pessimistic and optimistic cases).}
\label{figsfrdzcii}
\end{figure}

The models by \citet{madau14}, \citet{bethermin17}, and \citet{pillepich18} are also reported as black dashed, orange dot-dashed, and dark green solid curves, respectively. We notice that the more recent galaxy formation simulations 
(e.g. IllustrisTNG, \citealt{pillepich18}) and the model by \citet{bethermin17} are consistent with the ALPINE 1$\sigma$ lower error up to $z$$\sim$4 (excluding the $z$$=$3 bin, where our data are always higher), but significantly lower than the 
ALPINE $\rho_{\rm SFR}(z)$ at $z$$\gtrsim$4, with the difference becoming a factor of $\gtrsim$5--6 at $z$$\sim$6. 
The predictions by \citet{lagache18b} obtained by modelling the CIB are plotted as violet triple-dot-dashed curves
(showing both pessimistic and optimistic cases), and show good consistency with our data. 
In Figure~\ref{figsfrdzcii}, which is less crowded by data, we report the four models again for a better comparison with our results,
and we also add the prediction by \citet{manyiar18} obtained by modelling the CIB (light yellow dashed area). 
This latter prediction, similarly to the \citet{madau14} model, is lower than our data at $z$$\gtrsim$2.5.

From the comparison of all the data shown in Figure~\ref{figldz}, we can conclude that a significant amount of SF activity at high-$z$ is still missed by surveys sampling the UV rest frame of
galaxy emission. Indeed, all the far-IR/sub-mm and radio estimates agree within the uncertainties in showing an almost constant SFRD distribution from $z$$\sim$1 to 6, implying a
significant and increasing contribution of dust-obscured activity, which, starting from $z$$>$3--4, cannot be recovered by the dust-extinction corrected UV data. 
A similar discrepancy is observed with the predictions of galaxy formation simulations, that are not able to predict such a high amount of SFR in galaxies as we observe from $z$$\sim$3--4 to higher redshifts.
As discussed also by \citet{rowanrobinson16}, this result implies a significantly earlier start of the epoch of high SFRD than assumed by galaxy formation models and by UV-based works.
Therefore, the epoch of major activity in galaxies, corresponding to the rapid heavy element formation, extended at least from redshift 6 to redshift 1, which is at odds with the
predictions of the semi-analytic models for galaxy formation (e.g. \citealt{henriques15}), which set the epoch of intense star formation at $z$$\simeq$1--2. Our result strengthens the debated 
result of \citet{rowanrobinson16}, but also the previous conclusion of \citet{gruppioni15} showing that the semi-analytic models under-predict the high SFRs observed in the {\em Herschel} 
galaxies already at $z$$\gtrsim$1.5--2.  
%----------------------------------------------------------------- 
%----------------------------------------------------------------- 
%----------------------------------------------------------------- 
\begin{figure}
\includegraphics[width=8.8cm]{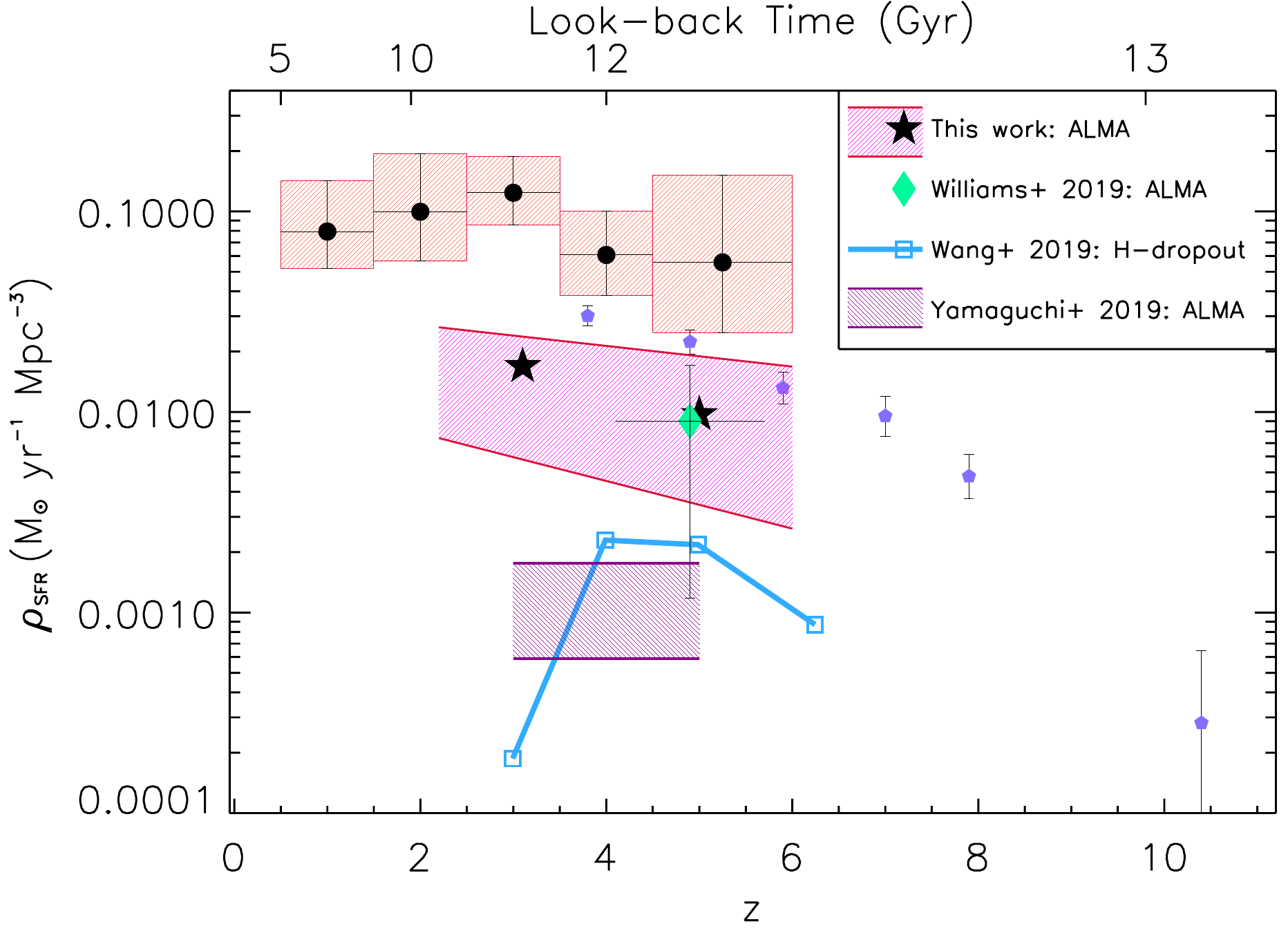}
\caption{Contribution of HST$+$near-IR dark galaxies to $\rho_{\rm SFR}(z)$: the derivation from this work (ALMA selection) in two redshift intervals (i.e. $z$$=$2.2--4.0 and 4.0--6.0) is shown via the magenta filled area and the black stars, and
is compared to the derivation by \citet{wang19} from H-dropout selection (blue open squares and line), by \citet{yamaguchi19} from ASAGAO (turquoise filled area), and
by \citet{williams19} from the ALMA selection (green diamond). 
The red boxes with black circles and the filled pentagons (the same as in Figure \ref{figldz}) represent the obscured and 
unobscured SFRD from this work and from \citet{bouwens15}, respectively. }
\label{figldz_dark}
\end{figure}
%----------------------------------------------------------------- 
%----------------------------------------------------------------- 
\subsection{Comparison with the SFRD derived from the ALPINE [C~II] luminosity functions}
\label{sec:ciisfrd}
In Figure~\ref{figsfrdzcii}, we compare the SFRD obtained by excluding the [C~II] emitters (red boxes and black circles: same as in Figure \ref{figldz}) to that 
estimated by integrating the best fit curve to the 4.5$<$$z$$<$6 LF obtained from all the continuum detections, including the two [C~II] emitters (yellow dashed boxes and brown open squares).
The inclusion of the two [C~II] emitters enhances our SFRD in the highest redshift bin,
causing a discontinuity with respect to the SFRD at lower redshifts,
due to the overdensity associated with the ALPINE targets to which these sources likely belong.
A similar -- and even more pronounced -- effect is observed with the [C~II] SFRD by
\citet{loiacono20} (also shown in the figure for comparison), where the SFRD derived for the detected [C~II] lines in the same
side band of the ALPINE targets (i.e. clustered:  green square and open triangle) 
is about an order of magnitude higher than the SFRD of the [C~II] emitters not 
associated with the targets (i.e. field: blue square and upside-down triangle). 
Therefore, we can conclude that by also including the sources detected because
they were associated with the primary targets of the ALMA observation, we would
have likely introduced a bias (overestimate) in our SFRD derivations.

%----------------------------------------------------------------- 
\subsection{Contribution of the optically and near-IR dark galaxies to the SFRD}
\label{sec:darksfrd}
In Figure~\ref{figldz_dark}, we show the estimated contribution to the SFRD at $z$$\simeq$3 and 5 from the ALPINE optically$+$near-IR dark galaxies. Our result, obtained by summing the SFR contribution of the HST$+$near-IR sources in the
two redshift intervals 2.2--4.0 and 4.0--6.0 (i.e. four and two sources, respectively, if we exclude the [C~II] emitters in the latter bin) 
is compared to previous estimates from the literature, obtained either through ALMA selection 
(e.g. \citealt{yamaguchi19, williams19}) or with different techniques (e.g. H-dropouts; \citealt{wang19}).
Despite the large uncertainties, our estimates are significantly higher than the SFRD contribution of the HST-dark galaxies selected by \citet{wang19} as H-dropouts, and of those selected
by \citet{yamaguchi19} from the ASAGAO ALMA survey. However, our result is consistent with the estimate based on a single sub-mm galaxy published by \citet{williams19} 
(i.e. $\rho_{\rm IR}^{\rm HSTdark}$($z$$\simeq$5)$=$0.9$^{+2.0}_{-0.7}$$\times$10$^{-2}$ M$_\odot$ yr$^{-1}$ Mpc$^{-3}$), while we find (1.5$\pm$0.9)$\times$10$^{-2}$ and (0.9$\pm$0.7)$\times$10$^{-2}$ M$_\odot$ yr$^{-1}$ Mpc$^{-3}$ at $z$$\simeq$3 and 5, respectively. From Figure~\ref{figldz_dark}, we note that the contribution to the co-moving SFRD of the HST$+$near-IR dark galaxies in ALPINE at $z$$\simeq$5 is almost equal 
to the extinction-corrected contribution from all the known UV-selected galaxies at similar redshifts. This means that the dust-obscured star formation also continues to contribute a significant fraction of the total
SFRD beyond $z$$>$3, and at least up to $z$$\simeq$6, where the available IR and mm estimates are still scanty.
Previous works, such as the detailed `super-deblending' analysis of {\em Herschel} fluxes in the GOODS-N field performed by \citet{liu18}, found that $z$$>$3 
dusty galaxies missed by optical-to-near-IR colour selection can significantly contribute to the SFRD at 3$<$$z$$<$6, and that far-IR$+$mm-derived SFRD are mostly independent of or complementary to those derived from optical and UV measurements.
We note, however, that the contribution at $z$$=$5 from the HST$+$near-IR dark galaxies is only $\sim$1/6 of the total (i.e. from all the sources), therefore the bulk of the difference between the corrected-UV and the total SFRD is not due to 
the dark galaxies, but more likely the dust correction of the UV samples that is too difficult to estimate from optical data.

The fact that we identified six dark galaxies in a survey of 24.9 arcmin$^2$ implies a source density of about 0.24 arcmin$^{-2}$, of the same order of that derived by 
\citet{williams19} (0.13 arcmin$^{-2}$), which is about a factor of three higher than the density of near-IR dark galaxies in ASAGAO (e.g. two sources in 26 arcmin$^2$: $\sim$8$\times$10$^{-2}$ arcmin$^{-2}$;
\citealt{yamaguchi19}). At $z$$>$3, our dark galaxies have a density of $\sim$0.12 arcmin$^{-2}$, which is $\sim$10$\times$ higher than that of $z$$>$4 SMGs (1--2$\times$10$^{-2}$ arcmin$^{-2}$; e.g. \citealt{danielson17,marrone18}). 
Similar densities, such as (0.042$\pm$0.028) arcmin$^{-2}$, are reported by \citet{riechers20} for optically dark CO emitters at $z$$>$5 detected down to an equivalent
870-$\mu$m flux density of $\sim$5 mJy.

By considering the volumes corresponding to each source in our survey, we derive a space density of HST$+$near-IR dark galaxies in ALPINE of $\sim$(1.2$\pm$0.7)$\times$10$^{-4}$ and 
(5.0$\pm$3.8)$\times$10$^{-5}$ Mpc$^{-3}$ at $z$$\simeq$3 and 5, respectively. The value found in the highest redshift interval is 
higher (though consistent within the uncertainties) than the source density of dark galaxies 
estimated by \citet{williams19} at $z$$\simeq$4.1--5.7 (2.9$\times$10$^{-5}$ Mpc$^{-3}$) and by \citet{riechers20} at $z$$>$5 (i.e. (1.0$\pm$0.7) $\times$10$^{-5}$ Mpc$^{-3}$).

%----------------------------------------------------------------- 
\subsection{Contribution of the optically and near-IR dark galaxies to the stellar mass density}
\label{sec:SMD}
Although the mass estimates for the ALPINE-detected, HST$+$near-R dark galaxies are very uncertain (given the paucity of photometric points available), these sources are likely to contribute 
significantly to the cosmic stellar mass density (SMD or $\rho*$) at high redshifts.
Indeed, by summing up the volume-weighted masses of our dark galaxies, we find that they might represent a fraction of the 
total SMD (as derived by \citealt{davidzon17} for the COSMOS15 galaxies) as high as $\sim$20\%, and $>$50\% at $z$$\simeq$3 and 5, respectively. They could even dominate the high-mass end of the stellar mass function at $z$$>$3 (see also \citealt{rodighiero07}). 
In fact, we find that the number density of the ALPINE dark galaxies with $M^*$$>$$10^{10.7}$ M$_{\odot}$ is $\sim$(5.1$\pm$3.7)$\times$10$^{-5}$ Mpc$^{-3}$, 
which is comparable to that of the more massive quiescent galaxies at $z$$\gsimeq$3--4 (e.g. $\sim$2$\times$10$^{-5}$ Mpc$^{-3}$; \citealt{gobat12,straatman14,song16,glazebrook17}).
The early formation of such a large number of massive, dusty galaxies is not predicted by the current
semi-analytical models (e.g. \citealt{henriques15}) and hydrodynamic simulations (e.g. \citealt{pillepich18}), which largely
underestimate the density of massive galaxies at high redshifts (see e.g. \citealt{alcade19,wang19}). 
Similarly, the galaxy formation models and simulations are also not able to explain the observed large density of IR luminous galaxies 
at $z$$>$2 (e.g. \citealt{gruppioni15,rowanrobinson16}).
The direct implication of these large abundances of massive and IR luminous (dusty) galaxies in the early Universe (not predicted by the up-to-date
state-of-the-art models) is that our current knowledge of the formation and evolution of massive/luminous galaxies is still far from being complete,
and  the relative theories might need important revisions.

In the near future, further investigations of the nature and physical parameters of the HST$+$near-IR dark galaxies will be necessary to
consolidate our results and conclusions. In particular, follow-up studies in the mid-IR (photometry and/or spectroscopy) with the James Webb Space Telescope, 
and in the sub-mm/mm (continuum and/or spectral-scanning) with ALMA or NOEMA will be the foreseen key observations. 
%----------------------------------------------------------------- 
%----------------------------------------------------------------- 
\section{Conclusions}
\label{sec:concl}
We used the 56 sources blindly detected in continuum (ALMA band 7, i.e. at 860 or 1000 $\mu$m) within the ALPINE survey to investigate the nature, evolution, and main properties 
of the dusty galaxy population across the 0.5$\lesssim$$z$$\lesssim$6 redshift range. The main points of our work can be summarised as follows:
   \begin{enumerate}
   \item We performed a detailed identification analysis, either by matching the positions of the ALPINE continuum sources with the available multi-wavelength and photo-$z$ catalogues, 
   or by looking for counterparts in the deep photometric images, then performing ad-hoc photometry and deriving photometric redshifts. Six of the continuum sources showed a
   faint counterpart only in the mid-IR, with no HST or near-IR matches. Five (two with no counterparts at all) have been identified with [C~II] emitters at $z$$\sim$5 (same $z$ as the ALPINE targets
   at the centre of the pointings).  For four sources, no counterpart was found in any of the available catalogues or images, though two are detected at significantly high SNR (6.7 and 9.3).
       \item We fully characterised the multi-wavelength SEDs of the ALPINE non-target sources by performing a detailed SED-fitting analysis and comparison with known template libraries of IR populations. The
       SED-fitting analysis provided the main physical parameters of the sources, which are $L_{\rm IR}$, SFR, $M^*$, galaxy class, 
       k-correction and, if needed, a photo-$z$ estimate. The median redshift of the
       whole ALPINE non-target, continuum-detected sub-mm galaxy population is $\overline{z}$$\simeq$2.84$\pm$0.18 (2.53$\pm$0.17 if we exclude the five sources at the same $z$ of the ALPINE targets), while for the HST and near-IR dark galaxies it is 
       $\overline{z}_{\rm dark}$$\simeq$3.5$\pm$0.5 (although their $z$-distribution shows two peaks around $z$$\sim$3 and 5). The ALPINE continuum sources on average resulted to be massive galaxies, 
       with stellar masses in the 10$^{10}$--10$^{11}$ M$_{\odot}$ range ($\overline{M}^*$$\simeq$10$^{10.6}$ M$_{\odot}$ for the HST$+$near-IR dark galaxies), though not as extreme
       as what was found in previous works. % at $z$$\sim$3, $\simeq$10$^{11.2}$ M$_{\odot}$ at $z$$\sim$5).
       \item We computed the rest-frame LFs at 250\,$\mu$m in different redshift bins, from 0.5$<$$z$$<$1.5 up to 4.5$<$$z$$<$6 and compared them with the {\em Herschel} and SCUBA-2 LFs at 
       the same wavelength available in the literature. The ALPINE LF is almost complementary to the previous ones, the former mostly sampling the faint end, and the latter the bright end. In the common 
       redshift and luminosity range, our results are more consistent with the {\em Herschel} ones.
      \item We integrated the SEDs over $\lambda_{\rm rest}$$=$8--1000\,$\mu$m, computed the total IR LFs in different redshift intervals (from $z$$\simeq$0.5 up to $z$$\sim$6), and studied its evolution
      with $z$. Although ALPINE mostly covers the faint end of the LFs, the global shape appears flat, with a low faint-end slope and a high bright end, not dropping at bright $L_{\rm IR}$. 
      There are no signs of a significant decrease in the normalisation nor of a change in shape from $z$$=$0.5 to $z$$=$6. Our results are in very good agreement with those from CO LFs by
      \citet{riechers19} and \citet{decarli16, decarli20}.
       \item We derived the co-moving SFRD over the $z$$\simeq$0.5--6 redshift range and the contribution of HST and near-IR dark galaxies at $z$$\simeq$3 and 5. The SFRD shows a flat distribution over the whole $z$-range, 
       with no significant decrease beyond the cosmic noon ($z$$\simeq$1--3). Our result is in agreement with those from previous far-IR and radio surveys, but higher than that found by optical/UV surveys at $z$$>$3. 
       The difference between our results and the optical/UV ones increases with redshift, reaching a factor of about 10 at $z$$\sim$6. The HST$+$near-IR dark galaxies contribute a significant fraction (about 17\%) of the total SFRD at high-$z$ ($>$3).
       We can conclude that a considerable amount of SF activity at high-$z$ is still missed by surveys sampling the UV rest frame (most of it not due to dark galaxies), with a significant and increasing contribution of dust-obscured 
       activity that cannot be recovered even by correcting the UV data for dust extinction. Similarly, the current galaxy formation models and simulations are not able to predict such a high amount 
       of SFR in dusty galaxies as is observed beyond cosmic noon.
        \item We derived the contribution of the ALPINE HST$+$near-IR dark galaxies to the cosmic mass density, notably finding that the number density of $M^*$$>$$10^{10.7}$ M$_{\odot}$ 
        dark galaxies is comparable to that of the more massive quiescent galaxies at $z$$\gtrsim$3. Given that neither the current
        semi-analytical models nor the more recent hydrodynamic simulations can explain the early formation of such a large number of massive dusty galaxies, we will need to revise our current understanding of the formation of massive/luminous galaxies.                  
        \end{enumerate}
%----------------------------------------------------------------- 
%----------------------------------------------------------------- 
\begin{acknowledgements}
We are grateful to the anonymous referee, whose constructive comments helped us to improve the manuscript.
CG acknowledges financial support from the Italian Space Agency under the ASI-INAF contract n. 2018-31-HH.0.
CG, FP, MT, AC, GR acknowledge the support from grant PRIN MIUR 2017$-$20173ML3WW$\_$001. 
GL acknowledges support from the European Research Council (ERC) under the European Union's Horizon 2020 
research and innovation programme (project CONCERTO, grant agreement No 788212) and from the Excellence 
Initiative of Aix-Marseille University-A*Midex, a French "Investissements d'Avenir" programme.
SF acknowledge support from the European Research Council (ERC) Consolidator Grant funding scheme 
(project ConTExt, grant No. 648179). The Cosmic Dawn Center is funded by the Danish National Research 
Foundation under grant No. 140. D.R. acknowledges support from the National Science Foundation under grant numbers 
AST-1614213 and AST-1910107. D.R. also acknowledges support from the Alexander von Humboldt Foundation through a 
Humboldt Research Fellowship for Experienced Researchers.
This work is based on observations taken by the 3D-HST Treasury Program (GO 12177 and 12328) 
with the NASA/ESA HST, which is operated by the Association of Universities for Research in Astronomy, 
Inc., under NASA contract NAS5$-$26555. Based on data products from observations made with ESO 
Telescopes at the La Silla Paranal Observatory under ESO programme ID 179.A-2005 and on data 
products produced by CALET and the Cambridge Astronomy Survey Unit on behalf of the UltraVISTA consortium.
This paper is dedicated to the memory of Olivier Le F\`evre, PI of the ALPINE survey.
\end{acknowledgements}

  \bibliographystyle{aa} % style aa.bst
  \bibliography{mybibliography} % your references Yourfile.bib

\end{document}